\documentclass[print, %% change to preprint for no formatting
               notlongauthorlist %% if affiliations take up more than a page use longauthorlist
              ]{nsr}

\usepackage{amsmath, bm}
\usepackage{physics}
\newcommand{\Li}{\rm Li}

\volume{00}

\artnum{00}

\firstpage{1}

\datesubmitted{XX Month Year}
\datereceived{XX Month Year} 
\daterevised{XX Month Year}
\dateaccepted{XX Month Year} 
\datepublished{XX Month Year}

\doinum{doi/number}

\copyrightyear{2026}

\author[1]{Amaresh Jaiswal\footnote{Corresponding author e-mail address: \tt a.jaiswal@niser.ac.in}$^{\orcidlink{0000-0001-5692-9167}}$}

\affil[1]{School of Physical Sciences, National Institute of Science Education and Research, An OCC of Homi Bhabha National Institute, Jatni-752050, India}

\runauth{Amaresh Jaiswal et al.}

\title{Spin-hydrodynamics in graphene and thermovortical magnetization}

\begin{document}

%% Start line numbering with \begin{linenumbers}, end it with \end{linenumbers}.
%% Or switch it on for the whole article with \linenumbers.
%\linenumbers

\maketitle

%%%%%%%%%%%%%%%%%%%%%%%%%%%%%%%%%%%%%%%%%%%%%%%%%%%%%%%%%%%%%%%%%%%%%%%%%

%% Insert abstract of the article
\begin{abstract}
\noindent
We examine the framework of relativistic spin hydrodynamics in the context of the Dirac quasiparticle fluid in graphene. We develop a spin-hydrodynamic model for a $(2+1)$-dimensional system of fermions under the condition of small spin polarization. Our analysis confirms that thermal vorticity, which satisfies the global equilibrium condition, is also a solution to the spin-hydrodynamic equations. Additionally, we calculate the magnetization of the system in global equilibrium and introduce a novel phenomenon—\emph{thermovortical magnetization}—resulting from thermal vorticity, which can be experimentally observed in graphene.
\end{abstract}

%% Write your article keywords

%%%%%%%%%%%%%%%%%%%%%%%%%%%%%%%%%%%%%%%%%%%%%%%%%%%%%%%%%%%%%

\section{Introduction}
\label{sec:intro}
% TODO: write your article here.
The hydrodynamic behavior of electrons in solid-state systems has attracted considerable attention in recent years from researchers across condensed-matter physics, hydrodynamics, and high-energy theory. A prominent platform where electron hydrodynamics has been extensively investigated is graphene~\cite{Ku:2019lgj, varnavides2020electron, PhysRevB.103.125106, PhysRevB.103.235152, PhysRevB.103.155128, PhysRevResearch.3.013290, DiSante:2019zrd, PhysRevB.103.115402, sulpizio2019visualizing, doi:10.1126/science.aau0685, ella2019simultaneous, bandurin2018fluidity, doi:10.1126/science.aad0201, Aung:2023vrr, Idrisov:2023}; see Refs.~\cite{eHD1, eHD2, eHD3} for recent reviews. Owing to its simple band structure and high carrier mobility, graphene provides an exceptionally rich setting for exploring collective electron transport. Graphene consists of a two-dimensional honeycomb lattice of carbon atoms, in which the low-energy quasiparticles obey an effective relativistic dispersion relation, $E(\mathbf{p})=\pm\, v_{\rm F}\,|\mathbf{p}|$, where $v_{\rm F}$ is the Fermi velocity of graphene~\cite{Wallace:1947qeg}. This linear dispersion relation, characteristic of massless Dirac fermions, was experimentally confirmed in Refs.~\cite{Novoselov:2005kj, Zhang:2005zz}. Moreover, Coulomb interactions between charge carriers in graphene can be sufficiently strong to drive the system into a hydrodynamic regime~\cite{eHD1}. As a result, graphene provides an ideal platform for testing theoretical ideas from relativistic hydrodynamics with experimental results, in a condensed-matter setting.

Relativistic hydrodynamics finds applications in fields such as astrophysics, cosmology and high-energy physics. Evolution of strongly interacting, hot and dense QCD matter, formed in high-energy heavy-ion collisions, has been modeled quite successfully within the framework of relativistic hydrodynamics~\cite{Heinz:2013th}. The recent observation of spin polarization in non-central heavy-ion collisions has prompted the development of relativistic spin-hydrodynamics, which enforces the conservation of total angular momentum in addition to the usual energy-momentum and net particle current conservation~\cite{Florkowski:2017ruc, Florkowski:2018fap, Bhadury:2020cop, Bhadury:2020puc, Becattini:2020ngo, Hu:2021pwh, Shi:2020htn, Fu:2020oxj, Speranza:2020ilk, Speranza:2021bxf, She:2021lhe, Peng:2021ago, Wang:2021ngp, Yi:2021unq, Wang:2021wqq, Florkowski:2019qdp, Singh:2020rht, Singh:2021man, Tinti:2016bav, Jaiswal:2012qm, Florkowski:2021wvk, Das:2022azr, Montenegro:2017rbu, Montenegro:2017lvf, Montenegro:2018bcf, Montenegro:2020paq, Gallegos:2021bzp, Hongo:2021ona, Cartwright:2021qpp, Gallegos:2022jow}. However, the fireball produced in heavy-ion collisions is extremely short-lived and certain key results of relativistic spin-hydrodynamics have yet to be conclusively verified through experiments.

One of the important predictions of relativistic spin-hydrodynamics is the existence of thermal vorticity as the global equilibrium solution~\cite{Becattini:2009wh, Becattini:2013fla, Florkowski:2017ruc}, which has been quite successful in explaining the global spin-polarization observed in relativistic heavy-ion collisions~\cite{Becattini:2016gvu, Karpenko:2016jyx, Li:2017slc}. However, attempts to apply thermal vorticity solution to explain longitudinal spin-polarization in relativistic heavy-ion collisions leads to opposite sign compared to the experimental results~\cite{Becattini:2017gcx, Florkowski:2019voj, Fu:2021pok, Becattini:2021suc, Becattini:2021iol, Florkowski:2021xvy}. This may be attributed to the fact that due to the short lifetime of the fireball created in relativistic heavy-ion collisions, the system does not have sufficient time to attain global equilibrium for spin dynamics in the transverse plane~\cite{Kapusta:2019sad, Ayala:2020ndx, Kumar:2023ghs, Hidaka:2023oze, Wagner:2024fhf, Banerjee:2024xnd}. On the other hand, such issues related to short lifetimes do not occur in the case of electron hydrodynamics in graphene, making it an ideal platform to validate the global equilibrium solution of relativistic spin hydrodynamics.

Near charge neutrality, graphene is predicted to exhibit behavior characteristic of a quantum-critical relativistic plasma, known as the "Dirac fluid," in which thermally excited electrons and holes with linear dispersion undergo frequent collisions~\cite{RevModPhys.81.109, Crossno2016, Lucas2016}. At charge neutrality, the quantum-critical scattering rate indicative of the Dirac fluid was experimentally observed in Ref.~\cite{Gallagher2019}. Further, with increased doping, emergence of two distinct current-carrying modes were also observed, one with zero total momentum and the other with nonzero total momentum, revealing a hallmark of relativistic hydrodynamics~\cite{Gallagher2019}. Therefore in graphene, electrons and holes form a plasma governed by relativistic hydrodynamic equations, analogous to those describing plasmas of hot quarks and gluons~\cite{Lucas2019}.

We emphasize that the relativistic notation employed throughout this work refers to the emergent Dirac structure of the low-energy quasiparticle theory of graphene and does not imply exact Lorentz invariance of the underlying condensed-matter system. In particular, the Fermi velocity $v_F$ replaces the speed of light as the characteristic velocity of the continuum Dirac theory, while the underlying lattice, electromagnetic environment, and material-dependent interactions explicitly break the fundamental Lorentz symmetry. We therefore use the covariant $(2+1)$-dimensional notation as a convenient formulation of the effective Dirac theory rather than as a statement about the microscopic symmetry of graphene. Our analysis is further restricted to the Dirac-fluid regime near charge neutrality, where $|\mu|/T$ is not parametrically large. The degenerate regime $\mu\gg T$, in which a different low-temperature hydrodynamic description is appropriate, is beyond the scope of the present work.

Before proceeding, it is important to clarify the physical meaning and scope of ``spin hydrodynamics" in the present context. In relativistic theories, spin is not a strictly conserved quantity, and its local density and current are known to depend on the choice of pseudo-gauge for the energy–momentum and spin tensors. Consequently, the interpretation of spin as an independent hydrodynamic variable is subtle and, in general, requires additional assumptions. In this work, we adopt the viewpoint that spin hydrodynamics provides an effective, coarse-grained description of slowly relaxing spin polarization, rather than a microscopic statement about exact spin conservation. Moreover, in the present work the spin variable appearing in relativistic spin hydrodynamics should be interpreted as the internal angular momentum associated with the Dirac quasiparticle representation describing the low-energy excitations of graphene. Therefore, it represents the internal angular-momentum degree of freedom appearing in the effective relativistic kinetic theory of Dirac quasiparticles.

In graphene, electronic spin is not conserved due to coupling to the lattice, spin–orbit interactions, and spin–phonon and impurity scattering processes. Nevertheless, a hydrodynamic description of spin may become meaningful in regimes where the spin relaxation time is parametrically longer than the electron–electron scattering time that governs charge and momentum equilibration~\cite{PhysRevLett.112.116602, HAN2012369, VanTuan2016}. In such a situation, spin polarization can be treated as an approximately conserved, long-lived degree of freedom on hydrodynamic time scales, analogous to other quasi-hydrodynamic variables encountered in electron fluids. Throughout this work, we restrict ourselves to this idealized but controlled regime.

In this article, we explore the framework of relativistic spin-hydrodynamics within the context of Dirac quasiparticle fluid in graphene. We develop the framework of spin-hydrodynamics for a $(2+1)$-dimensional system of fermions in the limit of small spin-polarization. The present study focuses on the non-dissipative limit of relativistic spin hydrodynamics and on global equilibrium configurations characterized by thermal vorticity. Our goal is not to establish the universal validity of spin hydrodynamics in graphene, but rather to explore its concrete and testable consequences if such a description applies. 

We re-affirm that thermal vorticity, which satisfies global equilibrium condition, is also a solution of the spin-hydrodynamic equations in $(2+1)$-dimensions. In particular, we show that global equilibrium with thermal vorticity leads to a vorticity-induced magnetization, which constitutes a genuine electromagnetic response and is therefore directly observable. This provides a physically transparent way to assess the relevance of spin-hydrodynamic concepts in a realistic condensed-matter system. We identify this novel phenomenon as \emph{thermovortical magnetization} and discuss its observable significance. We use the convention $g_{\mu\nu} = \hbox{diag}(+1,-1,-1)$ for the metric tensor. We employ $A\cdot B\equiv A_\mu B^\mu$ and $A:B\equiv A_{\mu\nu}B^{\mu\nu}$ to denote scalar products. Throughout the text we use natural units where $v_{\rm F} = \hbar = k_B =1$.

%%%%%%%%%%%%%%%%%%%%%%%%%%%%%%%%%%%%%%%%%%%%%%%%%%%%%%%%%%%%%

\section{Density operator and equilibrium}
\label{sec:den_op}

Relativistic spin hydrodynamics asserts that a complete description of a relativistic fluid, consisting of particles with intrinsic spin, necessitates the inclusion of a rank-3 spin tensor which is the expectation value of the tensor operator $\widehat{S}^{\lambda,\mu\nu}$. This spin-tensor operator (anti-symmetric in the last two indices) contributes to the total angular momentum operator as
\begin{equation}\label{J_op}
    \widehat{J}^{\lambda,\mu\nu} = x^\mu \, \widehat{T}^{\lambda\nu} - x^\nu \, \widehat{T}^{\lambda\mu} + \widehat{S}^{\lambda,\mu\nu},
\end{equation}
where $\widehat{T}^{\mu\nu}$ is the energy-momentum tensor operator. 

Within the framework of quantum statistical mechanics, local thermodynamic equilibrium is defined as the state that maximizes the entropy $S=-{\rm Tr}(\widehat{\rho}\ln\widehat{\rho})$, where $\widehat{\rho}$ is the density operator. Moreover, entropy is maximized subject to the constraint that the mean values of charge, energy, momentum, and spin densities remain fixed, leading to~\cite{Becattini:2014yxa, Becattini:2018duy, Florkowski:2018fap}
\begin{equation}\label{rho_LE}
    \widehat{\rho}_{\rm LE} = \frac{1}{Z_{\rm LE}} \exp\!\left[ -\!\!\int_\sigma \!\!d\sigma_\mu \!\left(\! \widehat{T}^{\mu\nu}\beta_\nu - \frac{1}{2}\omega_{\alpha\beta}\widehat{S}^{\mu,\alpha\beta} - \alpha \widehat{N}^\mu \!\right)\! \right].
\end{equation}
Here, $\sigma$ is a space-like hypersurface with element $d\sigma_\mu$, $Z_{\rm LE}$ is the local equilibrium partition function, $\widehat{N}^\mu$ is the conserved net particle number operator, and $\beta_\mu$, $\omega_{\alpha\beta}$, $\alpha$ are the relevant Lagrange multipliers. Using Eq.~\eqref{J_op}, the integral in Eq.~\eqref{rho_LE} can be expressed in terms of the conserved angular momentum operator as~\cite{Florkowski:2018fap}
\begin{equation}\label{I_LE}
\widehat{\cal{I}}_{\rm LE} = \int_\sigma \! d\sigma_\mu \left[ \widehat{T}^{\mu\nu} \big( \beta_\nu - \omega_{\nu\gamma}\, x^\gamma \big) - \frac{1}{2}\omega_{\alpha\beta}\widehat{J}^{\mu,\alpha\beta} - \alpha \widehat{N}^\mu \right].
\end{equation}
It is important to note that the conservation of particle current, energy-momentum and angular momentum implies that $\partial_\mu \widehat{N}^\mu = \partial_\mu \widehat{T}^{\mu\nu} = \partial_\lambda \widehat{J}^{\lambda,\mu\nu}=0$. Further, using expression of $\widehat{J}^{\lambda,\mu\nu}$ from Eq.~\eqref{J_op}, angular momentum conservation leads to $\partial_\lambda \widehat{S}^{\lambda,\mu\nu}=\widehat{T}^{\nu \mu} -\widehat{T}^{\mu \nu}$. 

In global equilibrium, the integral in Eq.~\eqref{I_LE} denoted as $\widehat{\cal{I}}_{\rm GE}$, must be independent of the choice of space-like hypersurface $\sigma$, ensuring that the density operator remains constant. This is achieved by requiring that the divergence of the integrand in Eq.~\eqref{I_LE} vanishes. Considering the conservation equations, the global equilibrium constraint imposes three conditions on the Lagrange multipliers
\begin{equation}\label{geq_cond}
\partial_\mu \beta_\nu = \omega_{\nu\mu}, \quad \partial_\mu \omega_{\nu\gamma} = 0, \quad \partial_\mu\alpha = 0.
\end{equation}
The above conditions implies that 
\begin{equation}\label{geq_sol}
\beta_\nu = b_\nu + \omega_{\nu\gamma}\, x^\gamma, \quad \omega_{\nu\gamma} = {\rm const.}, \quad \alpha = {\rm const.},
\end{equation}
where, $b_\nu$ is constant. Keeping in mind that $\omega_{\mu\nu}$ is anti-symmetric, the first condition in Eq.~\eqref{geq_cond} implies that $\beta_\nu$ should satisfy the Killing equation $\partial_\mu \beta_\nu + \partial_\nu \beta_\mu =0$. Moreover, Eq.~\eqref{geq_sol} implies that 
\begin{equation}\label{th_vort}
\omega_{\mu\nu} = -\frac{1}{2} \big( \partial_\mu\beta_\nu - \partial_\nu\beta_\mu \big) = {\rm const.},
\end{equation}
which is referred to as the thermal vorticity. In order to distinguish $\omega_{\mu\nu}$ from its global equilibrium value, thermal vorticity is often denoted by $\varpi_{\mu\nu}\equiv-\frac{1}{2} \big( \partial_\mu\beta_\nu - \partial_\nu\beta_\mu \big)$.

At this point, we note that the conservation equations $\partial_\mu \widehat{T}^{\mu\nu} = 0$ and $\partial_\lambda \widehat{S}^{\lambda,\mu\nu}=\widehat{T}^{\nu \mu} -\widehat{T}^{\mu \nu}$ are invariant under the so called \textit{pseudogauge freedom}. This symmetry reflects the non-uniqueness in the definitions of the energy–momentum and spin tensor operators, and is implemented through the following two transformations~\cite{Hehl:1976vr, Becattini:2011ev, Becattini:2025twu}
\begin{align}
\widehat{T}^{\prime \mu \nu} &= \widehat{T}^{\mu\nu} + \frac{1}  {2} \partial_\lambda \left( \widehat{\Phi}^{\lambda, \mu \nu} + \widehat{\Phi}^{\nu, \mu \lambda} + \widehat{\Phi}^{\mu, \nu \lambda} \right), \label{eq:PG1} \\
\widehat{S}^{\prime \lambda, \mu \nu} &= \widehat{S}^{\lambda, \mu \nu} - \widehat{\Phi}^{\lambda, \mu \nu} . \label{eq:PG2}
\end{align}
Here the tensor operator $\widehat{\Phi}^{\lambda, \mu \nu}$ is referred to as the superpotentials and is antisymmetric under the exchange of its last two indices, $\widehat{\Phi}^{\lambda, \mu \nu} = -\,\widehat{\Phi}^{\lambda, \nu \mu}$. The transformations given in Eqs.~\eqref{eq:PG1} and \eqref{eq:PG2}, when applied to Eq.~\eqref{I_LE} in global equilibrium with appropriate boundary conditions, leads to the change~\cite{Becattini:2025twu}
\begin{equation}\label{I_GE}
\delta\widehat{\cal{I}}_{\rm GE} =  \int_\sigma \! d\sigma_\lambda \left[ \frac{1}{2}\, \widehat{\Phi}^{\lambda, \mu \nu} \left\{ \omega_{\mu\nu} + \frac{1}{2} \left( \partial_\mu \beta_\nu - \partial_\nu \beta_\mu \right) \right\} - \, \widehat{\Phi}^{\mu, \lambda \nu} \left\{ \frac{1}{2} \left(  \partial_\mu \beta_\nu + \partial_\nu \beta_\mu \right) \right\} \right].
\end{equation}
Using the global equilibrium conditions for Lagrange multipliers given in Eqs.~\eqref{geq_cond} - \eqref{th_vort}, we see that the change in density operator via $\delta\widehat{\cal{I}}_{\rm GE}$ vanishes. This demonstrates that the global-equilibrium observables, which are the primary focus of this work, are invariant under pseudo-gauge transformations.

%%%%%%%%%%%%%%%%%%%%%%%%%%%%%%%%%%%%%%%%%%%%%%%%%%%%%%%%%%%%%

\section{Relativistic spin-hydrodynamics}
\label{sec:spin_hyd}

In order to formulate relativistic spin-hydrodynamics, applicable to $2+1$-dimensional system of fermions, we start with single particle, phase-space distribution functions for particles and anti-particles~\cite{Becattini:2013fla, Florkowski:2017ruc}. To account for spin degrees of freedom, the phase-space distribution functions at space-time position $x$ and momentum $p$ are generalized to spin density matrices
\begin{align}
    f^+_{rs}(x,p) &= \frac{1}{2m}\, \bar{u}_r(p)\, X^+ \, u_s(p), \label{fprs}\\
    f^-_{rs}(x,p) &= -\frac{1}{2m}\, \bar{v}_s(p)\, X^-\, v_r(p),  \label{fmrs}
\end{align}
for particles and anti-particles, respectively. In the above equations, $m$ is the mass of the particles and $u_r(p)$, $v_r(p)$ are bispinors with the normalization $\bar{u}_r(p)\, u_s(p) = 2m\,\delta_{rs}$ and $\bar{v}_r(p)\, v_s(p)=-2 m\,\delta_{rs}$\footnote{These normalization conditions ensure that the massless limit of Eqs.~\eqref{fprs} and \eqref{fmrs} are well defined.}. Here the spin indices $r$ and $s$ can take values of 1 and 2.

Adopting the notation used in Refs.~\cite{Becattini:2013fla, Florkowski:2017ruc}, we introduce the matrices
\begin{equation}\label{Xpm}
    X^{\pm} = \left[ \exp\left( \beta\cdot p \,\mp\, \alpha \right) \exp\left( \pm \frac{1}{2} \omega:\Sigma \right) + I \right]^{-1}, 
\end{equation}
where $\beta^\mu$, $\alpha$ and $\omega^{\mu\nu}$ are the Lagrange multipliers, as introduced in Eq.~\eqref{rho_LE}, and $p^\mu=(|{\boldsymbol p}|, {\boldsymbol p})$. In terms of physical quantities, we have $\alpha = \mu/T$ and $\beta^\mu = u^\mu/T$, with $T$, $\mu$ and $u^\mu$ being the temperature, chemical potential and fluid velocity, respectively, where $u^\mu$ is normalized to $u\cdot u=1$. Further, the Lagrange multiplier $\omega^{\mu\nu}$ is termed as the polarization tensor and $\Sigma^{\mu\nu}$ is the spin operator which can be represented using the Dirac gamma matrices as $\Sigma^{\mu\nu} = (i/4) [\gamma^\mu,\gamma^\nu]$. It is important to note that the Dirac gamma matrices satisfy the relations of the Clifford algebra $\{\gamma^\mu,\gamma^\nu\}\equiv\gamma^\mu\gamma^\nu+\gamma^\nu\gamma^\mu=2\,g^{\mu\nu}$. Further, we consider $4\times 4$ matrix representation of Dirac Gamma matrices with the Greek indices $(\mu,\nu,\cdots)$ taking values of $0, 1, 2$. 

At this point, it is important to clarify the interpretation of the spin degree of freedom appearing in the present formulation. In graphene, the low-energy electronic excitations are described by an effective Dirac theory in which the two-component structure of the quasiparticle wavefunction originates from the sublattice degree of freedom and is commonly referred to as pseudospin. In the present work, the spin tensor entering the relativistic spin-hydrodynamic framework should be interpreted as the hydrodynamic current associated with the internal angular momentum of the effective Dirac quasiparticles. The corresponding single-particle internal angular momentum is generated by the Lorentz generators $\Sigma^{\mu\nu}=\frac{i}{4}[\gamma^\mu,\gamma^\nu]$, which act on the internal spinor space arising from the sublattice (and, in the four-component formulation, valley) degrees of freedom. Consequently, $\Sigma^{\mu\nu}$ is constructed from the same underlying internal degrees of freedom that give rise to the pseudospin structure of graphene quasiparticles, while serving as the generator of Lorentz transformations within the effective Dirac theory. The present work therefore investigates the consequences of applying the relativistic spin-hydrodynamic framework to this effective internal angular momentum of Dirac quasiparticles, rather than to the microscopic electron spin.

The fundamental conserved hydrodynamic quantities can be obtained by employing the distribution functions in Eqs.~\eqref{fprs} and \eqref{fmrs}. For a $(2+1)$-dimensional system of massless fermions, the charge current and energy-momentum tensor can be expressed as~\cite{DeGroot:1980dk}
\begin{align}
    N^\mu &= \int \frac{d^2p}{2 (2\pi)^2 |{\boldsymbol p}|}\,  p^\mu \left[ \tr( X^+ ) - \tr ( X^- )  \right], \label{Nmu}\\
    T^{\mu\nu} &= \int \frac{d^2p}{2 (2\pi)^2 |{\boldsymbol p}|}  \,  p^\mu \,  p^\nu \left[ \tr( X^+ ) +  \tr ( X^- )  \right]. \label{Tmunu}
\end{align}
For the spin-tensor, we use the phenomenological form~\cite{Becattini:2009wh}
\begin{equation}\label{Slmunu}
    S^{\lambda,\mu\nu} = \int \frac{d^2p}{2 (2\pi)^2 |{\boldsymbol p}|}\,  p^\lambda \tr\left[ ( X^+  -  X^- ) \Sigma^{\mu\nu} \right],
\end{equation}
which is consistent with the global equilibrium solution of thermal vorticity~\cite{Florkowski:2017ruc}. It is important to note that the pseudospin associated with the sublattice degree of freedom is helically locked to the momentum in the massless Dirac Hamiltonian and is therefore not an independent degree of freedom that can be separately polarized~\cite{RevModPhys.81.109}. However, the hydrodynamic spin tensor should not be interpreted as the polarization of the microscopic lattice pseudospin. Instead, it represents the expectation value of the internal Dirac angular momentum appearing in the effective kinetic-theory formulation of relativistic spin hydrodynamics.

In order to evaluate the trace and perform the integrals in Eqs.~\eqref{Nmu}, \eqref{Tmunu} and \eqref{Slmunu}, we consider that the spin-polarization induced in the medium is small. In the limit of small polarization, i.e., for small values of $\omega^{\mu\nu}$, Eq.~\eqref{Xpm} can be Taylor expanded up to linear order in $\omega^{\mu\nu}$ as
\begin{equation}\label{Xpmlin}
    X^{\pm} = f_0^\pm\,I \pm \frac{1}{2} \left( \omega:\Sigma \right) f_0^\pm\,\tilde{f_0}^\pm,
\end{equation}
where, $f_0^\pm\equiv [\exp(\beta\cdot p\mp\alpha)+1]^{-1}$ and $\tilde{f_0}^\pm\equiv 1-f_0^\pm$. The trace of operators appearing in Eqs.~\eqref{Nmu}, \eqref{Tmunu} and \eqref{Slmunu} can be evaluated by noting that $\tr(\Sigma^{\mu\nu})=0$ and $\tr(\Sigma^{\mu\nu}\Sigma^{\alpha\beta})=g^{\mu\alpha}g^{\nu\beta}-g^{\mu\beta}g^{\nu\alpha}$. Performing the momentum integration, we obtain
\begin{align}
    N^\mu &= n\,u^\mu, \label{Nmu_hyd}\\
    T^{\mu\nu} &= (\varepsilon+P) u^\mu u^\nu - P g^{\mu\nu}, \label{Tmunu_hyd}\\
    S^{\lambda,\mu\nu} &= w\, u^\lambda \omega^{\mu\nu}, \label{Slmunu_hyd}
\end{align}
where, $n$, $\varepsilon$, $P$ and $w$ are net number density, energy density, pressure and spin density, respectively. These hydrodynamic quantities are obtained as
\begin{align}
    n &= \frac{T^2}{\pi} \Big[ \Li_2\left( - {\rm e}^{-\alpha} \right) - \Li_2\left( - {\rm e}^{\alpha} \right) \Big], \label{n_den}\\
    \varepsilon &= \frac{2\,T^3}{\pi} \Big[ - \Li_3\left( - {\rm e}^{-\alpha} \right) - \Li_3\left( - {\rm e}^{\alpha} \right) \Big], \label{e_den}\\
    P &= \frac{\varepsilon}{2}, \qquad w = \frac{T^2}{4\pi}\ln\left( 2+ 2\cosh\alpha \right), \label{P_w}
\end{align}
where $\Li_s(z)$ is the polylogarithmic function of order $s$ and argument $z$. The formalism developed here allows for a finite ratio $\alpha=\mu/T$, although our physical interpretation in terms of a quantum-critical Dirac fluid and the proposed experimental realization is focused on the near-neutrality regime. In this regime, the chemical potential may remain finite and spatially varying while satisfying the equilibrium condition of constant $\mu/T$. We therefore do not consider the strongly degenerate limit $\mu/T\gg1$, where the relevant hydrodynamic degrees of freedom and transport coefficients may differ from those considered here.

One may also introduce a magnetization tensor associated with the internal angular momentum of the quasiparticle ensemble as \cite{suttorp1970covariant, vanWeert1970, Bhadury:2022ulr}
\begin{equation}\label{magnetization}
M^{\mu\nu} = m\int  \frac{d^2p}{2 (2\pi)^2 |{\boldsymbol p}|} \tr\left[ ( X^+  -  X^- )\, m^{\mu\nu} \right].
\end{equation}
Following the effective relativistic kinetic-theory description, we adopt the phenomenological identification $m^{\alpha\beta}=\chi\, \Sigma^{\alpha\beta}$ for the magnetic dipole moment operator $m^{\alpha\beta}$. The present work does not attempt a microscopic derivation of this relation from the underlying tight-binding model of graphene; rather, it investigates the consequences of this effective coupling within the relativistic spin-hydrodynamic framework. Further, we employ the conventional choice for effective coupling parameter, $\chi=\frac{g\,e}{2\,m}$, which reproduces the standard gyromagnetic structure appearing in relativistic kinetic theory~\cite{suttorp1970covariant, Weickgenannt:2019dks}. Here, $g$ represents the g-factor, $e$ is the elementary charge\footnote{Not to be confused with Napier's constant/Euler's Number, ${\rm e}$.} and $m$ is the mass. In the context of graphene this parameter should be interpreted as an effective coupling describing the electromagnetic response of the Dirac quasiparticles, rather than as the microscopic magnetic moment of the lattice electrons. Note that the 1/$m$ in the expression for $\chi$ cancels with the $m$ in the definition of magnetization tensor, Eq.~\eqref{magnetization}, resulting in a well defined massless limit. Evaluating the trace and performing the momentum integration in Eq.~\eqref{magnetization}, we obtain
\begin{equation}\label{mag_hyd}
    M^{\mu\nu} = \frac{g\,e}{8\,\pi}\, T\,\omega^{\mu\nu},
\end{equation}
which is one of the main results of the present work.

The thermovortical magnetization discussed here is distinct from the conventional spin Hall effect, although both involve spin- or angular-momentum-related responses in Dirac materials. The spin Hall effect generally refers to a nonequilibrium transverse spin-current response generated by an applied electric field and mediated by spin-orbit coupling~\cite{RevModPhys.87.1213}. In graphene, such responses are associated with intrinsic, Rashba, or proximity-induced spin-orbit interactions~\cite{PhysRevLett.95.226801}. In contrast, the effect considered in the present work is an equilibrium magnetization induced by thermal vorticity, $M^{\mu\nu} \propto T\omega^{\mu\nu}$, and does not require an applied electric field or spin-orbit coupling as its leading source. Moreover, the spin degree of freedom employed here represents the internal angular momentum of the effective Dirac quasiparticle rather than the microscopic electron spin. Thus, thermovortical magnetization should be viewed as a distinct equilibrium rotational response rather than as a manifestation of the conventional spin Hall effect.

%%%%%%%%%%%%%%%%%%%%%%%%%%%%%%%%%%%%%%%%%%%%%%%%%%%%%%%%%%%%%

\section{Conservation laws and hydrodynamic equations}
\label{sec:cons_laws}

Conservation of net particle number, energy and momentum can be expressed as
\begin{equation}\label{N_Tmn_cons}
    \partial_\mu N^\mu = 0, \qquad \partial_\mu T^{\mu\nu} = 0.
\end{equation}
Moreover, conservation of angular momentum implies $\partial_\lambda J^{\lambda, \mu\nu}=0$, where $J^{\lambda, \mu\nu} = x^\mu T^{\lambda\nu} - x^\nu T^{\lambda\mu} + S^{\lambda, \mu\nu}$. Since energy-momentum tensor in the present work is a symmetric tensor, see Eqs.~\eqref{Tmunu} and \eqref{Tmunu_hyd}, the conservation of angular momentum requires that the spin tensor $S^{\lambda, \mu\nu}$ must also satisfy the conservation equation~\cite{Hehl:1976vr},
\begin{equation}\label{Slmn_cons}
    \partial_\lambda S^{\lambda, \mu\nu} = 0.
\end{equation}
Note that Eqs.~\eqref{N_Tmn_cons} and \eqref{Slmn_cons} constitute equations of motion for the hydrodynamic variables $T$, $\mu$, $u^\mu$ and $\omega^{\mu\nu}$.

Using the expressions for the conserved quantities, Eqs.~\eqref{Nmu_hyd}-\eqref{Slmunu_hyd} in the conservation equations, Eqs.~\eqref{N_Tmn_cons} and \eqref{Slmn_cons}, we obtain the equations of motion
\begin{align}
    \dot{n} + n\,\theta &= 0, \label{ndot}\\
    \dot{\varepsilon} + (\varepsilon+P)\,\theta &= 0, \label{edot}\\
    (\varepsilon+P)\,\dot{u^\alpha} - (g^{\alpha\beta}-u^\alpha u^\beta)\, \partial_\beta P &= 0, \label{udot}\\
    w\,\dot{\omega}^{\mu\nu} + \dot{w}\,\omega^{\mu\nu} + w\,\theta\,\omega^{\mu\nu} &= 0, \label{omegadot}
\end{align}
where, $\dot{A}\equiv u^\mu\partial_\mu A$ represents co-moving derivative and $\theta\equiv\partial_\mu u^\mu$ is the expansion scalar. Note that Eqs.~\eqref{edot} and \eqref{udot} follow from projection of energy-momentum conservation equation, along and orthogonal to $u^\mu$, respectively. In the following, we consider the global equilibrium condition of thermal vorticity, Eq.~\eqref{th_vort}, and verify that it is indeed a solution of Eqs.~\eqref{ndot}-\eqref{omegadot}.

%%%%%%%%%%%%%%%%%%%%%%%%%%%%%%%%%%%%%%%%%%%%%%%%%%%%%%%%%%%%%

\section{Thermal vorticity solution}
\label{sec:th_vort}

In order to express thermal vorticity in terms of hydrodynamic variables, we consider rigid vortical motion of the fluid in the two dimensional space. In this case, components of the hydrodynamic flow velocity, $u^\mu=(u^0,u^1,u^2)$, is defined as
\begin{equation}\label{u_vort}
    u^0 = \gamma, \quad u^1 = - \, \gamma \, \Omega \, y, \quad u^2 = \gamma \, \Omega \, x,
\end{equation}
where $\Omega$ is a positive constant indicating rotation speed, $\gamma \equiv 1/\sqrt{1 - \Omega^2 r^2}$ is the local Lorentz-$\gamma$ factor, and $r \equiv \sqrt{x^2 + y^2}$ represents the distance from the center of the vortex. The profile for temperature and chemical potential is determined as
\begin{equation}\label{T_mu_vort}
    T = T_0 \, \gamma, \quad \mu = \mu_0 \, \gamma
\end{equation}
with $T_0$ and $\mu_0$ being constants. Using the definition $\beta^\mu=u^\mu/T$ in Eq.~\eqref{th_vort}, we obtain
\begin{equation}\label{omega_vort}
    \omega_{\mu\nu} = \varpi_{\mu\nu} = \left( \begin{array}{ccc}
0 & 0   & 0   \\
0 & 0   & \Omega/T_0  \\
0 & -\Omega/T_0 & 0   \\
\end{array} 
\right), 
\end{equation}
which is a constant matrix. Note that the thermal-vorticity solution corresponds to a rigidly rotating fluid configuration, which preserves rotational symmetry about the rotation axis.

In order to verify that Eqs.~\eqref{u_vort}-\eqref{omega_vort} satisfies the hydrodynamic equations of motion, Eqs.~\eqref{ndot}-\eqref{omegadot}, we begin by noting that the comoving derivative takes the form
\begin{equation}\label{umu_delmu}
    u^\mu\partial_\mu = -\gamma\,\Omega\left( y\frac{\partial}{\partial x} - x\frac{\partial}{\partial y} \right).
\end{equation}
Keeping in mind that $\alpha=\mu/T=\mu_0/T_0$ is a constant, and using the above expression for the comoving derivative, we obtain
\begin{equation}\label{dots_vort}
    \dot{n} = 0, \quad \dot{\varepsilon}  = 0, \quad \dot{P} = 0, \quad \dot{w} = 0,  \quad \theta = 0 .
\end{equation}
Further, using the equation-of-state $\varepsilon=2\,P$, the scaled derivative of pressure can be shown to be equal to the comoving derivative of fluid velocity, i.e., 
\begin{equation}\label{vec_vort}
    \left(\frac{1}{\varepsilon+P}\right) \partial^\alpha P = \dot{u}^\alpha =  - \gamma^2 \Omega^2 (0,\, x,\, y).
\end{equation}
It is now apparent from Eqs.~\eqref{dots_vort} and \eqref{vec_vort} that Eqs.~\eqref{u_vort}-\eqref{omega_vort} represents a solution of the hydrodynamic equations of motion, i.e., Eqs.~\eqref{ndot}-\eqref{omegadot}. 

It is important to emphasize that in this section we have verified that the thermal-vorticity configuration given by Eqs.~\eqref{u_vort}--\eqref{omega_vort} satisfies the hydrodynamic equations of motion, Eqs.~\eqref{ndot}--\eqref{omegadot}, rather than deriving this solution from a specified set of initial conditions. Indeed, it follows directly from Eq.~\eqref{omegadot} that if the initial value of $\omega^{\mu\nu}$ vanishes, it remains zero throughout the evolution within the ideal hydrodynamic framework. On the other hand, a nonvanishing initial value of $\omega^{\mu\nu}$ can lead to a vortical configuration consistent with the thermal-vorticity solution. In the physical setup considered here, such conditions may arise through localized Joule heating generated by nanowires, as described in the next section. In this setup, the Joule heating generated in the nanowires acts as an external source term in the hydrodynamic conservation equations, Eqs.~\eqref{N_Tmn_cons} and \eqref{Slmn_cons}, leading to spatially inhomogeneous energy deposition in the fluid at early times. Once the external heating is switched off, this inhomogeneous energy distribution serves as the initial condition for the subsequent hydrodynamic evolution. The resulting spatially nonuniform temperature gradients in the fluid drive pressure-induced flows whose velocity field need not remain irrotational. Consequently, the hydrodynamic evolution can generate a nonzero $\omega^{\mu\nu}$, allowing the system to approach a vortical global equilibrium configuration characterized by thermal vorticity $\varpi^{\mu\nu}$.

%%%%%%%%%%%%%%%%%%%%%%%%%%%%%%%%%%%%%%%%%%%%%%%%%%%%%%%%%%%%%

\section{Thermovortical magnetization}
\label{sec:thvort_mag}

To understand the observable consequence of the spin hydrodynamics framework on Dirac quasiparticle fluid in graphene, we consider magnetization with thermal vorticity solution in global equilibrium. Using Eq.~\eqref{omega_vort} in Eq.~\eqref{mag_hyd}, we obtain
\begin{equation}\label{ther_vort_mag}
    M_{\mu\nu} = \frac{g\,e}{8\,\pi}\,\frac{\Omega}{\sqrt{1 - \Omega^2 r^2}}
    \left( \begin{array}{ccc}
0 & 0   & 0   \\
0 & 0   & 1  \\
0 & -1 & 0   \\
\end{array} 
\right).
\end{equation}
The above relation suggests that the condition of global equilibrium with thermal vorticity leads to a nonzero magnetization of the quasiparticle fluid. The magnetization obtained in the above equation should be interpreted within the effective relativistic description employed in this work. This effect can be interpreted as a relativistic analogue of the Barnett effect observed in solids~\cite{Barnett:1935wyv}, in which mechanical rotation generates an effective magnetization through the coupling between rotation and internal angular momentum. In the present framework this coupling arises from the thermal-vorticity term appearing in the local equilibrium density operator. Consequently, Eq.~\eqref{ther_vort_mag} should be viewed as the equilibrium magnetic response of a Dirac quasiparticle fluid rather than as a direct microscopic spin polarization of lattice electrons. Further, our calculation suggests that the \emph{thermovortical magnetization} in Eq.~\eqref{ther_vort_mag} increases as a function of radial distance from the center of the vortex. 

%--------------------------------
\begin{figure}[t!]
    \centering
    \includegraphics[width=0.6\linewidth]{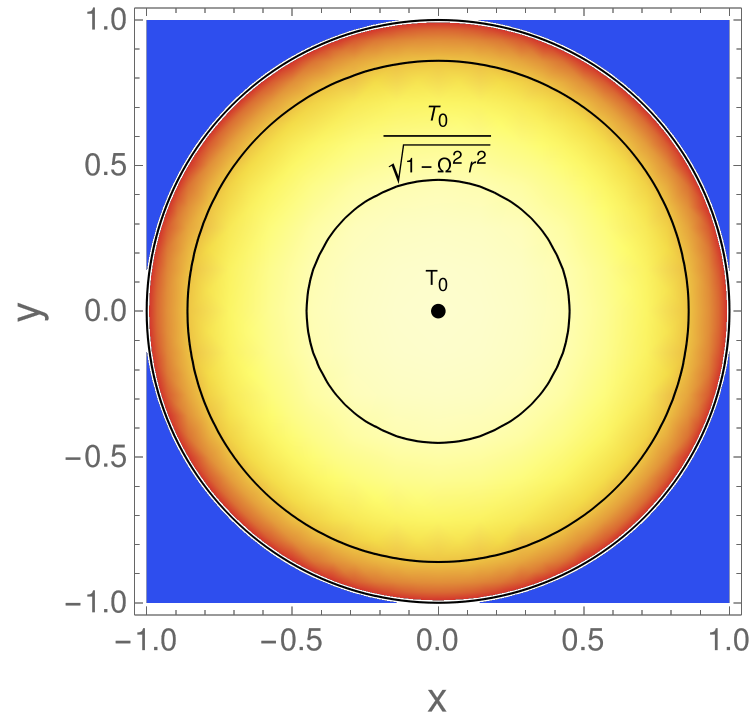}
    \hspace{-0.2cm}
    \raisebox{0.8cm}{\includegraphics[width=0.093\linewidth]{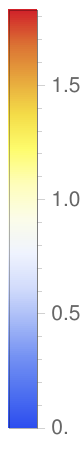}}
    \vspace{-0.3cm}
    \caption{Schematic representation of temperature gradient on the surface of graphene, shown as a disc of radius 1 unit. The temperature at the center is $T_0$ and the temperature on the surface of graphene increases outwards as a function of radius $r$, as shown in the figure. The black concentric circles represent nanowires used to maintain the specific temperature profile radially outwards; see text for details.}
    \vspace*{-0.2cm}
    \label{fig_1}
\end{figure}
%--------------------------------

It is important to note that in order to experimentally observe thermovortical magnetization in graphene, the system has to be initialized in such a way that it relaxes to the thermal vorticity state of global equilibrium. We propose that such an initial condition can be achieved by providing a gradient to the thermodynamic quantities given in Eq.~\eqref{T_mu_vort}. For instance, an external source may be applied to maintain a radially increasing temperature gradient $T=T_0/\sqrt{1 - \Omega^2 r^2}$; as shown in Fig.~\ref{fig_1}. Joule heating in nanowires can serve as an external heat source to maintain the desired temperature gradient on graphene substrate~\cite{10.1063/1.1847714, PhysRevLett.98.187202, PhysRevB.84.054437}. Once the external source maintaining the temperature gradient is switched off, the subsequent evolution is expected to relax toward a vortical equilibrium configuration characterized by thermal vorticity. This thermal vorticity state eventually leads to thermovortical magnetization, as given in Eq.~\eqref{ther_vort_mag}. We emphasize that, while the magnetization induced by thermal vorticity may be small, it is nonetheless a measurable phenomenon, as described below.

Firstly, we note that the expression for magnetization in Eq.~\eqref{ther_vort_mag} represents the \emph{magnetization density}. Consequently, the total magnitude of magnetization is given by $M=(ge/8\pi)\,\Omega\,\gamma\,(\pi R^2)$, where $R$ is the radius of the graphene sample. The central temperature of the graphene sheet, $T_0$, as illustrated in Fig.~\ref{fig_1}, can be maintained at approximately 80\,K using a liquid nitrogen cold finger. On the other hand, Joule heating in nanowires can readily raise the temperature to about 400\,K at the edges of the sample $( r = R$)~\cite{PhysRevB.84.054437, RAVICHANDRAN201514}. Using Eq.~\eqref{T_mu_vort}, these values corresponds to a Lorentz gamma factor of $\gamma = T / T_0 \approx 5$, indicating a sizable dimensionless flow velocity in the effective Dirac description. Moreover, this leads to $\Omega R \approx v_{\mathrm{F}} \approx 10^6\,\mathrm{m/s}$, the typical Fermi velocity in graphene. Substituting the above values and fundamental constants into the expression for $M$, we obtain $M \approx 10^{-13} \, R ~ \text{[A$\cdot$m}^2\text{]}$, where $R$ is in meters. The minimum magnetization detectable with current technology is of the order of $\sim 10^{-17} \, \mathrm{A\cdot m}^2$~\cite{10.1063/5.0259803}. Therefore, to observe the thermovortical magnetization experimentally, the graphene sample must have a radius $R \sim 100\,\mu\mathrm{m}$, which is typical size of graphene sheets.

To further elaborate on the proposed experimental setup, we emphasize that Fig.~\ref{fig_1} serves only as a conceptual schematic. The concentric nanowires illustrated in the figure can be fabricated on a substrate using established techniques~\cite{doi:10.1073/pnas.0914031107}, upon which a hexagonal boron nitride (hBN)-encapsulated graphene sample may be transferred. The thermal characteristics of the underlying substrate will play a critical role in maintaining the desired radial temperature gradient. Given the high thermal conductivity of graphene, the lattice temperature is expected to rapidly equilibrate with that of the substrate. Additionally, electron-phonon interactions will drive the electronic temperature toward the lattice temperature, thereby establishing the intended temperature profile. However, important practical considerations remain, including heat dissipation at the graphene--substrate interface~\cite{Xu_2012} and the precise electrical interfacing of the nanowires. These factors must be carefully addressed in any experimental realization of the proposed setup.

To assess how closely the initial temperature profile must be in order to relax to $T=T_0\gamma$ exact configuration, it would likely require detailed numerical modeling of heat transport and hydrodynamic behavior within the graphene sample. Nevertheless, we note that there are two distinct classes of global equilibrium solutions in this context: one characterized by spatially uniform temperature, and another associated with thermal vorticity. When a radially increasing temperature profile is provided as the initial condition, the system is expected to evolve toward the nearest equilibrium state, which would correspond to a thermal vorticity configuration of the form $T'=T_0\gamma'$, possibly associated with a modified angular velocity $\Omega'$. Although this state may not quantitatively match the specific profile used in our calculations, it would still give rise to a nonzero magnetization. Thus, even if the final configuration deviates quantitatively from the idealized setup, our analysis provides a robust qualitative prediction for the emergence of magnetization driven by thermal vorticity. We emphasize that the purpose of the proposed experimental setup is to illustrate a possible physical realization of such a vortical equilibrium state rather than to provide a detailed dynamical mechanism for its formation.

%%%%%%%%%%%%%%%%%%%%%%%%%%%%%%%%%%%%%%%%%%%%%%%%%%%%%%%%%%%%%

\section{Summary and outlook}
\label{sec:sum_out}

In this work, the framework of relativistic spin-hydrodynamics was explored within the context of Dirac quasiparticle fluid in graphene. A spin-hydrodynamic model for a $(2+1)$-dimensional system of fermions was developed under the assumption of small spin polarization. It was confirmed that thermal vorticity, which satisfies the global equilibrium condition, also serves as a solution to the spin-hydrodynamic equations in $(2+1)$-dimensions. Furthermore, the magnetization of the system in global equilibrium was computed, leading to the introduction of a new phenomenon, \emph{thermovortical magnetization}, which results from thermal vorticity and can be observed experimentally in graphene.

A few remarks are in order regarding the interpretation and limitations of the present results. The formulation employed here is based on ideal (non-dissipative) relativistic spin hydrodynamics and assumes small spin polarization. In realistic graphene samples, spin polarization relaxes due to spin–orbit coupling and interactions with the lattice, and these effects are not captured at the level of the present analysis. Consequently, the results should be understood as describing the leading-order behavior in a regime where spin relaxation is sufficiently slow compared to the characteristic hydrodynamic time scales.

The thermovortical magnetization derived in this work is a global-equilibrium result, tied to the existence of a Killing-vector solution for the inverse-temperature field. Away from global equilibrium, dissipative corrections and relaxation effects are expected to modify the magnitude and spatial profile of the magnetization. However, the existence of a vorticity-induced magnetization itself is expected to be robust, as it follows from general equilibrium considerations and corresponds to a physical electromagnetic response. As such, it is not an artifact of a particular pseudo-gauge choice, as demonstrated in Sec.~\ref{sec:den_op}, even though intermediate definitions of spin density and spin current may depend on that choice.

From a broader perspective, the main result of this work is to identify a concrete observable consequence of relativistic spin hydrodynamics in a condensed-matter system. Whether spin hydrodynamics emerges as an accurate effective theory in graphene ultimately depends on material-specific relaxation mechanisms and experimental conditions. The present analysis therefore provides a benchmark equilibrium prediction that can guide both future theoretical extensions as well as experimental efforts aimed at probing rotational and vorticity-induced spin phenomena in electron fluids.

This work thus proposes a viable route toward applying and testing predictions of relativistic spin hydrodynamics in graphene. While our analysis is restricted to a non-dissipative formulation, it is of considerable interest to extend the framework to include dissipative effects, such as rotational viscosity and boost heat conductivity~\cite{Hattori:2019lfp}, in the context of relativistic spin-hydrodynamics in graphene. Incorporating such effects requires determining the associated transport coefficients from the underlying microscopic theory of interacting Dirac quasiparticles. In addition, the influence of external magnetic fields can be naturally incorporated within the framework of spin magnetohydrodynamics~\cite{Bhadury:2022ulr}. We leave a systematic investigation of these directions for future work.

A generalized formulation of the current framework can be developed to describe transport phenomena in systems such as Weyl semimetals. Topological Weyl semimetals exhibit linear dispersion near specific points in momentum space, known as Dirac or Weyl nodes, serving as three-dimensional analogs of graphene. Weyl semimetals exhibit unique transport phenomena arising from their low-energy excitations, which behave as chiral, massless fermions~\cite{Hosur:2013kxa, Vafek:2013mpa, Yan:2016euz, Burkov:2017rgl, Armitage:2017cjs}. These materials provide an intriguing platform to apply and extend relativistic spin-hydrodynamics due to the Lorentz-invariant structure of their effective field theory, combined with the presence of strong spin-orbit coupling, Berry curvature effects, and topological responses such as the chiral magnetic and chiral vortical effects. We leave the exploration of these promising directions to future investigations.

%%%%%%%%%%%%%%%%%%%%%%%%%%%%%%%%%%%%%%%%%%%%%%%%%%%%%%%%%%%%%

\section*{Acknowledgments}

The author expresses gratitude to Hiranmaya Mishra, Sabyasachi Ghosh, and Sourav Dey for valuable discussions. The author gratefully acknowledges the warm hospitality of IIT Gandhinagar where part of this work was completed.

%%%%%%%%%%%%%%%%%%%

\bigskip\noindent
{\bf Funding:} This research was funded by Department of Atomic Energy (DAE), India.

%%%%%%%%%%%%%%%%%%%

\bigskip\noindent
{\bf Conflicts of Interest:} The author declares no conflicts of interest.

%%%%%%%%%%%%%%%%%%%%%%%%%%%%%%%%%%%%%%%%%%%%%%%%%%%%%%%%%%%%%%%%%%%%%%%%%

%% References with BibTeX:
%%
\bibliographystyle{nsr}
\bibliography{ref}

@article{Montenegro:2018bcf,
    author = "Montenegro, David and Torrieri, Giorgio",
    title = "{Causality and dissipation in relativistic polarizable fluids}",
    eprint = "1807.02796",
    archivePrefix = "arXiv",
    primaryClass = "hep-th",
    doi = "10.1103/PhysRevD.100.056011",
    journal = "Phys. Rev. D",
    volume = "100",
    number = "5",
    pages = "056011",
    year = "2019"
}

@article{Becattini:2009wh,
    author = "Becattini, F. and Tinti, L.",
    title = "{The Ideal relativistic rotating gas as a perfect fluid with spin}",
    eprint = "0911.0864",
    archivePrefix = "arXiv",
    primaryClass = "gr-qc",
    doi = "10.1016/j.aop.2010.03.007",
    journal = "Annals Phys.",
    volume = "325",
    pages = "1566--1594",
    year = "2010"
}

@article{Bhadury:2020cop,
    author = "Bhadury, Samapan and Florkowski, Wojciech and Jaiswal, Amaresh and Kumar, Avdhesh and Ryblewski, Radoslaw",
    title = "{Dissipative Spin Dynamics in Relativistic Matter}",
    eprint = "2008.10976",
    archivePrefix = "arXiv",
    primaryClass = "nucl-th",
    doi = "10.1103/PhysRevD.103.014030",
    journal = "Phys. Rev. D",
    volume = "103",
    number = "1",
    pages = "014030",
    year = "2021"
}

@article{Bhadury:2020puc,
    author = "Bhadury, Samapan and Florkowski, Wojciech and Jaiswal, Amaresh and Kumar, Avdhesh and Ryblewski, Radoslaw",
    title = "{Relativistic dissipative spin dynamics in the relaxation time approximation}",
    eprint = "2002.03937",
    archivePrefix = "arXiv",
    primaryClass = "hep-ph",
    doi = "10.1016/j.physletb.2021.136096",
    journal = "Phys. Lett. B",
    volume = "814",
    pages = "136096",
    year = "2021"
}

@article{Becattini:2013fla,
    author = "Becattini, F. and Chandra, V. and Del Zanna, L. and Grossi, E.",
    title = "{Relativistic distribution function for particles with spin at local thermodynamical equilibrium}",
    eprint = "1303.3431",
    archivePrefix = "arXiv",
    primaryClass = "nucl-th",
    doi = "10.1016/j.aop.2013.07.004",
    journal = "Annals Phys.",
    volume = "338",
    pages = "32--49",
    year = "2013"
}

@article{Florkowski:2019qdp,
    author = "Florkowski, Wojciech and Kumar, Avdhesh and Ryblewski, Radoslaw and Singh, Rajeev",
    title = "{Spin polarization evolution in a boost invariant hydrodynamical background}",
    eprint = "1901.09655",
    archivePrefix = "arXiv",
    primaryClass = "hep-ph",
    doi = "10.1103/PhysRevC.99.044910",
    journal = "Phys. Rev. C",
    volume = "99",
    number = "4",
    pages = "044910",
    year = "2019"
}

@article{Fu:2021pok,
    author = "Fu, Baochi and Liu, Shuai Y. F. and Pang, Longgang and Song, Huichao and Yin, Yi",
    title = "{Shear-Induced Spin Polarization in Heavy-Ion Collisions}",
    eprint = "2103.10403",
    archivePrefix = "arXiv",
    primaryClass = "hep-ph",
    doi = "10.1103/PhysRevLett.127.142301",
    journal = "Phys. Rev. Lett.",
    volume = "127",
    number = "14",
    pages = "142301",
    year = "2021"
}

@article{suttorp1970covariant,
  title={Covariant equations of motion, for a charged particle with a magnetic dipole moment},
  author={Suttorp, LG and De Groot, SR},
  journal={Il Nuovo Cimento A (1965-1970)},
  volume={65},
  number={1},
  pages={245--274},
  year={1970},
  publisher={Springer}
}

@article{Gallegos:2021bzp,
    author = {Gallegos, A. D. and G\"ursoy, U. and Yarom, A.},
    title = "{Hydrodynamics of spin currents}",
    eprint = "2101.04759",
    archivePrefix = "arXiv",
    primaryClass = "hep-th",
    doi = "10.21468/SciPostPhys.11.2.041",
    journal = "SciPost Phys.",
    volume = "11",
    pages = "041",
    year = "2021"
}

@article{Florkowski:2021wvk,
    author = "Florkowski, Wojciech and Ryblewski, Radoslaw and Singh, Rajeev and Sophys, Gabriel",
    title = "{Spin polarization dynamics in the non-boost-invariant background}",
    eprint = "2112.01856",
    archivePrefix = "arXiv",
    primaryClass = "hep-ph",
    doi = "10.1103/PhysRevD.105.054007",
    journal = "Phys. Rev. D",
    volume = "105",
    number = "5",
    pages = "054007",
    year = "2022"
}

@article{Becattini:2021suc,
    author = "Becattini, F. and Buzzegoli, M. and Palermo, A.",
    title = "{Spin-thermal shear coupling in a relativistic fluid}",
    eprint = "2103.10917",
    archivePrefix = "arXiv",
    primaryClass = "nucl-th",
    doi = "10.1016/j.physletb.2021.136519",
    journal = "Phys. Lett. B",
    volume = "820",
    pages = "136519",
    year = "2021"
}

@article{Fu:2020oxj,
    author = "Fu, Baochi and Xu, Kai and Huang, Xu-Guang and Song, Huichao",
    title = "{Hydrodynamic study of hyperon spin polarization in relativistic heavy ion collisions}",
    eprint = "2011.03740",
    archivePrefix = "arXiv",
    primaryClass = "nucl-th",
    doi = "10.1103/PhysRevC.103.024903",
    journal = "Phys. Rev. C",
    volume = "103",
    number = "2",
    pages = "024903",
    year = "2021"
}

@article{Singh:2021man,
    author = "Singh, Rajeev and Shokri, Masoud and Ryblewski, Radoslaw",
    title = "{Spin polarization dynamics in the Bjorken-expanding resistive MHD background}",
    eprint = "2103.02592",
    archivePrefix = "arXiv",
    primaryClass = "hep-ph",
    doi = "10.1103/PhysRevD.103.094034",
    journal = "Phys. Rev. D",
    volume = "103",
    number = "9",
    pages = "094034",
    year = "2021"
}

@article{She:2021lhe,
    author = "She, Duan and Huang, Anping and Hou, Defu and Liao, Jinfeng",
    title = "{Relativistic viscous hydrodynamics with angular momentum}",
    eprint = "2105.04060",
    archivePrefix = "arXiv",
    primaryClass = "nucl-th",
    doi = "10.1016/j.scib.2022.10.020",
    journal = "Sci. Bull.",
    volume = "67",
    pages = "2265--2268",
    year = "2022"
}

@article{Speranza:2021bxf,
    author = "Speranza, Enrico and Bemfica, Fabio S. and Disconzi, Marcelo M. and Noronha, Jorge",
    title = "{Challenges in solving chiral hydrodynamics}",
    eprint = "2104.02110",
    archivePrefix = "arXiv",
    primaryClass = "hep-th",
    doi = "10.1103/PhysRevD.107.054029",
    journal = "Phys. Rev. D",
    volume = "107",
    number = "5",
    pages = "054029",
    year = "2023"
}

@article{Becattini:2021iol,
    author = "Becattini, F. and Buzzegoli, M. and Inghirami, G. and Karpenko, I. and Palermo, A.",
    title = "{Local Polarization and Isothermal Local Equilibrium in Relativistic Heavy Ion Collisions}",
    eprint = "2103.14621",
    archivePrefix = "arXiv",
    primaryClass = "nucl-th",
    doi = "10.1103/PhysRevLett.127.272302",
    journal = "Phys. Rev. Lett.",
    volume = "127",
    number = "27",
    pages = "272302",
    year = "2021"
}

@article{Singh:2020rht,
    author = "Singh, Rajeev and Sophys, Gabriel and Ryblewski, Radoslaw",
    title = "{Spin polarization dynamics in the Gubser-expanding background}",
    eprint = "2011.14907",
    archivePrefix = "arXiv",
    primaryClass = "hep-ph",
    doi = "10.1103/PhysRevD.103.074024",
    journal = "Phys. Rev. D",
    volume = "103",
    number = "7",
    pages = "074024",
    year = "2021"
}

@article{Weickgenannt:2019dks,
    author = "Weickgenannt, Nora and Sheng, Xin-Li and Speranza, Enrico and Wang, Qun and Rischke, Dirk H.",
    title = "{Kinetic theory for massive spin-1/2 particles from the Wigner-function formalism}",
    eprint = "1902.06513",
    archivePrefix = "arXiv",
    primaryClass = "hep-ph",
    doi = "10.1103/PhysRevD.100.056018",
    journal = "Phys. Rev. D",
    volume = "100",
    number = "5",
    pages = "056018",
    year = "2019"
}

@article{Florkowski:2021xvy,
    author = "Florkowski, Wojciech and Kumar, Avdhesh and Mazeliauskas, Aleksas and Ryblewski, Radoslaw",
    title = "{Effect of thermal shear on longitudinal spin polarization in a thermal model}",
    eprint = "2112.02799",
    archivePrefix = "arXiv",
    primaryClass = "hep-ph",
    doi = "10.1103/PhysRevC.105.064901",
    journal = "Phys. Rev. C",
    volume = "105",
    number = "6",
    pages = "064901",
    year = "2022"
}

@article{Das:2022azr,
    author = "Das, Arpan and Florkowski, Wojciech and Kumar, Avdhesh and Ryblewski, Radoslaw and Singh, Rajeev",
    title = "{Semi-classical Kinetic Theory for Massive Spin-half Fermions with Leading-order Spin Effects}",
    eprint = "2203.15562",
    archivePrefix = "arXiv",
    primaryClass = "hep-th",
    reportNumber = "INT-PUB-23-032",
    doi = "10.5506/APhysPolB.54.8-A4",
    journal = "Acta Phys. Polon. B",
    volume = "54",
    number = "8",
    pages = "8--A4",
    year = "2023"
}

@article{Montenegro:2017rbu,
    author = "Montenegro, David and Tinti, Leonardo and Torrieri, Giorgio",
    title = "{Ideal relativistic fluid limit for a medium with polarization}",
    eprint = "1701.08263",
    archivePrefix = "arXiv",
    primaryClass = "hep-th",
    doi = "10.1103/PhysRevD.96.056012",
    journal = "Phys. Rev. D",
    volume = "96",
    number = "5",
    pages = "056012",
    year = "2017",
    note = "[Addendum: Phys.Rev.D 96, 079901 (2017)]"
}

@article{Montenegro:2017lvf,
    author = "Montenegro, David and Tinti, Leonardo and Torrieri, Giorgio",
    title = "{Sound waves and vortices in a polarized relativistic fluid}",
    eprint = "1703.03079",
    archivePrefix = "arXiv",
    primaryClass = "hep-th",
    doi = "10.1103/PhysRevD.96.076016",
    journal = "Phys. Rev. D",
    volume = "96",
    number = "7",
    pages = "076016",
    year = "2017"
}

@article{Montenegro:2020paq,
    author = "Montenegro, David and Torrieri, Giorgio",
    title = "{Linear response theory and effective action of relativistic hydrodynamics with spin}",
    eprint = "2004.10195",
    archivePrefix = "arXiv",
    primaryClass = "hep-th",
    doi = "10.1103/PhysRevD.102.036007",
    journal = "Phys. Rev. D",
    volume = "102",
    number = "3",
    pages = "036007",
    year = "2020"
}

@article{Gallegos:2022jow,
    author = "Gallegos, A. D. and Gursoy, U. and Yarom, A.",
    title = "{Hydrodynamics, spin currents and torsion}",
    eprint = "2203.05044",
    archivePrefix = "arXiv",
    primaryClass = "hep-th",
    doi = "10.1007/JHEP05(2023)139",
    journal = "JHEP",
    volume = "05",
    pages = "139",
    year = "2023"
}

@article{Hongo:2021ona,
    author = "Hongo, Masaru and Huang, Xu-Guang and Kaminski, Matthias and Stephanov, Mikhail and Yee, Ho-Ung",
    title = "{Relativistic spin hydrodynamics with torsion and linear response theory for spin relaxation}",
    eprint = "2107.14231",
    archivePrefix = "arXiv",
    primaryClass = "hep-th",
    reportNumber = "RIKEN-iTHEMS-Report-21",
    doi = "10.1007/JHEP11(2021)150",
    journal = "JHEP",
    volume = "11",
    pages = "150",
    year = "2021"
}

@article{Cartwright:2021qpp,
    author = "Cartwright, Casey and Amano, Markus Garbiso and Kaminski, Matthias and Noronha, Jorge and Speranza, Enrico",
    title = "{Convergence of hydrodynamics in a rotating strongly coupled plasma}",
    eprint = "2112.10781",
    archivePrefix = "arXiv",
    primaryClass = "hep-th",
    doi = "10.1103/PhysRevD.108.046014",
    journal = "Phys. Rev. D",
    volume = "108",
    number = "4",
    pages = "046014",
    year = "2023"
}

@article{Hu:2021pwh,
    author = "Hu, Jin",
    title = "{Relativistic first-order spin hydrodynamics via the Chapman-Enskog expansion}",
    eprint = "2111.03571",
    archivePrefix = "arXiv",
    primaryClass = "hep-ph",
    doi = "10.1103/PhysRevD.105.076009",
    journal = "Phys. Rev. D",
    volume = "105",
    number = "7",
    pages = "076009",
    year = "2022"
}

@article{Wang:2021wqq,
    author = "Wang, Dong-Lin and Xie, Xin-Qing and Fang, Shuo and Pu, Shi",
    title = "{Analytic solutions of relativistic dissipative spin hydrodynamics with radial expansion in Gubser flow}",
    eprint = "2112.15535",
    archivePrefix = "arXiv",
    primaryClass = "hep-ph",
    doi = "10.1103/PhysRevD.105.114050",
    journal = "Phys. Rev. D",
    volume = "105",
    number = "11",
    pages = "114050",
    year = "2022"
}

@article{Yi:2021unq,
    author = "Yi, Cong and Pu, Shi and Gao, Jian-Hua and Yang, Di-Lun",
    title = "{Hydrodynamic helicity polarization in relativistic heavy ion collisions}",
    eprint = "2112.15531",
    archivePrefix = "arXiv",
    primaryClass = "hep-ph",
    doi = "10.1103/PhysRevC.105.044911",
    journal = "Phys. Rev. C",
    volume = "105",
    number = "4",
    pages = "044911",
    year = "2022"
}

@article{Wang:2021ngp,
    author = "Wang, Dong-Lin and Fang, Shuo and Pu, Shi",
    title = "{Analytic solutions of relativistic dissipative spin hydrodynamics with Bjorken expansion}",
    eprint = "2107.11726",
    archivePrefix = "arXiv",
    primaryClass = "nucl-th",
    doi = "10.1103/PhysRevD.104.114043",
    journal = "Phys. Rev. D",
    volume = "104",
    number = "11",
    pages = "114043",
    year = "2021"
}

@article{Peng:2021ago,
    author = "Peng, Hao-Hao and Zhang, Jun-Jie and Sheng, Xin-Li and Wang, Qun",
    title = "{Ideal Spin Hydrodynamics from the Wigner Function Approach}",
    eprint = "2107.00448",
    archivePrefix = "arXiv",
    primaryClass = "hep-th",
    doi = "10.1088/0256-307X/38/11/116701",
    journal = "Chin. Phys. Lett.",
    volume = "38",
    number = "11",
    pages = "116701",
    year = "2021"
}

@article{Shi:2020htn,
    author = "Shi, Shuzhe and Gale, Charles and Jeon, Sangyong",
    title = "{From chiral kinetic theory to relativistic viscous spin hydrodynamics}",
    eprint = "2008.08618",
    archivePrefix = "arXiv",
    primaryClass = "nucl-th",
    doi = "10.1103/PhysRevC.103.044906",
    journal = "Phys. Rev. C",
    volume = "103",
    number = "4",
    pages = "044906",
    year = "2021"
}

@article{Li:2017slc,
    author = "Li, Hui and Pang, Long-Gang and Wang, Qun and Xia, Xiao-Liang",
    title = "{Global $\Lambda$ polarization in heavy-ion collisions from a transport model}",
    eprint = "1704.01507",
    archivePrefix = "arXiv",
    primaryClass = "nucl-th",
    doi = "10.1103/PhysRevC.96.054908",
    journal = "Phys. Rev. C",
    volume = "96",
    number = "5",
    pages = "054908",
    year = "2017"
}

@article{Hattori:2019lfp,
    author = "Hattori, Koichi and Hongo, Masaru and Huang, Xu-Guang and Matsuo, Mamoru and Taya, Hidetoshi",
    title = "{Fate of spin polarization in a relativistic fluid: An entropy-current analysis}",
    eprint = "1901.06615",
    archivePrefix = "arXiv",
    primaryClass = "hep-th",
    reportNumber = "RIKEN-iTHEMS-Report-19, YITP-19-15",
    doi = "10.1016/j.physletb.2019.05.040",
    journal = "Phys. Lett. B",
    volume = "795",
    pages = "100--106",
    year = "2019"
}

@article{Florkowski:2017ruc,
    author = "Florkowski, Wojciech and Friman, Bengt and Jaiswal, Amaresh and Speranza, Enrico",
    title = "{Relativistic fluid dynamics with spin}",
    eprint = "1705.00587",
    archivePrefix = "arXiv",
    primaryClass = "nucl-th",
    doi = "10.1103/PhysRevC.97.041901",
    journal = "Phys. Rev. C",
    volume = "97",
    number = "4",
    pages = "041901",
    year = "2018"
}

@article{Heinz:2013th,
    author = "Heinz, Ulrich and Snellings, Raimond",
    title = "{Collective flow and viscosity in relativistic heavy-ion collisions}",
    eprint = "1301.2826",
    archivePrefix = "arXiv",
    primaryClass = "nucl-th",
    doi = "10.1146/annurev-nucl-102212-170540",
    journal = "Ann. Rev. Nucl. Part. Sci.",
    volume = "63",
    pages = "123--151",
    year = "2013"
}

@article{Speranza:2020ilk,
    author = "Speranza, Enrico and Weickgenannt, Nora",
    title = "{Spin tensor and pseudo-gauges: from nuclear collisions to gravitational physics}",
    eprint = "2007.00138",
    archivePrefix = "arXiv",
    primaryClass = "nucl-th",
    doi = "10.1140/epja/s10050-021-00455-2",
    journal = "Eur. Phys. J. A",
    volume = "57",
    number = "5",
    pages = "155",
    year = "2021"
}

@article{Florkowski:2018fap,
    author = "Florkowski, Wojciech and Kumar, Avdhesh and Ryblewski, Radoslaw",
    title = "{Relativistic hydrodynamics for spin-polarized fluids}",
    eprint = "1811.04409",
    archivePrefix = "arXiv",
    primaryClass = "nucl-th",
    doi = "10.1016/j.ppnp.2019.07.001",
    journal = "Prog. Part. Nucl. Phys.",
    volume = "108",
    pages = "103709",
    year = "2019"
}

@article{Becattini:2020ngo,
    author = "Becattini, Francesco and Lisa, Michael A.",
    title = "{Polarization and Vorticity in the Quark\textendash{}Gluon Plasma}",
    eprint = "2003.03640",
    archivePrefix = "arXiv",
    primaryClass = "nucl-ex",
    doi = "10.1146/annurev-nucl-021920-095245",
    journal = "Ann. Rev. Nucl. Part. Sci.",
    volume = "70",
    pages = "395--423",
    year = "2020"
}

@book{DeGroot:1980dk,
  author    = {de Groot, S. R. and van Leeuwen, W. A. and van Weert, Ch. G.},
  title     = {Relativistic Kinetic Theory: Principles and Applications},
  publisher = {North-Holland Publishing Company},
  address   = {Amsterdam},
  year      = {1980},
  isbn      = {0444854533}
}

@article{Becattini:2016gvu,
    author = "Becattini, F. and Karpenko, I. and Lisa, M. and Upsal, I. and Voloshin, S.",
    title = "{Global hyperon polarization at local thermodynamic equilibrium with vorticity, magnetic field and feed-down}",
    eprint = "1610.02506",
    archivePrefix = "arXiv",
    primaryClass = "nucl-th",
    doi = "10.1103/PhysRevC.95.054902",
    journal = "Phys. Rev. C",
    volume = "95",
    number = "5",
    pages = "054902",
    year = "2017"
}

@phdthesis{vanWeert1970,
  author       = {van Weert, C. G.},
  title        = {On the Relativistic Kinetic Theory of Particles with a Magnetic Dipole Moment in an External Electromagnetic Field},
  school       = {University of Amsterdam},
  year         = {1970},
  address      = {Netherlands},
  type         = {PhD thesis}
}

@article{Ku:2019lgj,
    author = "Ku, Mark J. H. and others",
    title = "{Imaging viscous flow of the Dirac fluid in graphene}",
    doi = "10.1038/s41586-020-2507-2",
    journal = "Nature (London)",
    volume = "583",
    number = "7817",
    pages = "537--541",
    year = "2020"
}

@article{varnavides2020electron,
  title={Electron hydrodynamics in anisotropic materials},
  author={Varnavides, Georgios and Jermyn, Adam S and Anikeeva, Polina and Felser, Claudia and Narang, Prineha},
  journal={Nat. Commun.},
  volume={11},
  number={1},
  pages={4710},
  year={2020},
  publisher={Nature Publishing Group UK London},
  url={https://doi.org/10.1038/s41467-020-18553-y}
}

@article{PhysRevB.103.125106,
  title = {Electron hydrodynamics of two-dimensional anomalous Hall materials},
  author = {Hasdeo, Eddwi H. and Ekstr\"om, Johan and Idrisov, Edvin G. and Schmidt, Thomas L.},
  journal = {Phys. Rev. B},
  volume = {103},
  issue = {12},
  pages = {125106},
  numpages = {10},
  year = {2021},
  month = {Mar},
  publisher = {American Physical Society},
  doi = {10.1103/PhysRevB.103.125106},
  url = {https://link.aps.org/doi/10.1103/PhysRevB.103.125106}
}

@article{PhysRevB.103.235152,
  title = {Beyond Ohm's law: Bernoulli effect and streaming in electron hydrodynamics},
  author = {Hui, Aaron and Oganesyan, Vadim and Kim, Eun-Ah},
  journal = {Phys. Rev. B},
  volume = {103},
  issue = {23},
  pages = {235152},
  numpages = {13},
  year = {2021},
  month = {Jun},
  publisher = {American Physical Society},
  doi = {10.1103/PhysRevB.103.235152},
  url = {https://link.aps.org/doi/10.1103/PhysRevB.103.235152}
}

@article{PhysRevB.103.155128,
  title = {Electron-phonon hydrodynamics},
  author = {Huang, Xiaoyang and Lucas, Andrew},
  journal = {Phys. Rev. B},
  volume = {103},
  issue = {15},
  pages = {155128},
  numpages = {26},
  year = {2021},
  month = {Apr},
  publisher = {American Physical Society},
  doi = {10.1103/PhysRevB.103.155128},
  url = {https://link.aps.org/doi/10.1103/PhysRevB.103.155128}
}

@article{PhysRevResearch.3.013290,
  title = {Artificial electric field and electron hydrodynamics},
  author = {Tavakol, Omid and Kim, Yong Baek},
  journal = {Phys. Rev. Res.},
  volume = {3},
  issue = {1},
  pages = {013290},
  numpages = {14},
  year = {2021},
  month = {Mar},
  publisher = {American Physical Society},
  doi = {10.1103/PhysRevResearch.3.013290},
  url = {https://link.aps.org/doi/10.1103/PhysRevResearch.3.013290}
}

@article{DiSante:2019zrd,
    author = "Di Sante, Domenico and Erdmenger, Johanna and Greiter, Martin and Matthaiakakis, Ioannis and Meyer, Ren\'e and Rodr\'\i{}guez Fern\'andez, David and Thomale, Ronny and van Loon, Erik and Wehling, Tim",
    title = "{Turbulent hydrodynamics in strongly correlated Kagome metals}",
    
    doi = "10.1038/s41467-020-17663-x",
    journal = "Nat. Commun.",
    volume = "11",
    number = "1",
    pages = "3997",
    year = "2020"
}

@article{PhysRevB.103.115402,
  title = {Hydrodynamic collective modes in graphene},
  author = {Narozhny, B. N. and Gornyi, I. V. and Titov, M.},
  journal = {Phys. Rev. B},
  volume = {103},
  issue = {11},
  pages = {115402},
  numpages = {16},
  year = {2021},
  month = {Mar},
  publisher = {American Physical Society},
  doi = {10.1103/PhysRevB.103.115402},
  url = {https://link.aps.org/doi/10.1103/PhysRevB.103.115402}
}

@article{sulpizio2019visualizing,
  title={Visualizing Poiseuille flow of hydrodynamic electrons},
  author={Sulpizio, Joseph A and Ella, Lior and Rozen, Asaf and Birkbeck, John and Perello, David J and Dutta, Debarghya and Ben-Shalom, Moshe and Taniguchi, Takashi and Watanabe, Kenji and Holder, Tobias and others},
  journal={Nature (London)},
  volume={576},
  number={7785},
  pages={75--79},
  year={2019},
  publisher={Nature Publishing Group UK London},
  URL={https://doi.org/10.1038/s41586-019-1788-9}
}

@article{doi:10.1126/science.aau0685,
author = {A. I. Berdyugin  and S. G. Xu  and F. M. D. Pellegrino  and R. Krishna Kumar  and A. Principi  and I. Torre  and M. Ben Shalom  and T. Taniguchi  and K. Watanabe  and I. V. Grigorieva  and M. Polini  and A. K. Geim  and D. A. Bandurin },
title = {Measuring Hall viscosity of graphene's electron fluid},
journal = {Science},
volume = {364},
number = {6436},
pages = {162-165},
year = {2019},
doi = {10.1126/science.aau0685},
URL = {https://www.science.org/doi/abs/10.1126/science.aau0685},
}

@article{ella2019simultaneous,
  title={Simultaneous voltage and current density imaging of flowing electrons in two dimensions},
  author={Ella, Lior and Rozen, Asaf and Birkbeck, John and Ben-Shalom, Moshe and Perello, David and Zultak, Johanna and Taniguchi, Takashi and Watanabe, Kenji and Geim, Andre K and Ilani, Shahal and others},
  journal={Nat. Nanotechnol.},
  volume={14},
  number={5},
  pages={480--487},
  year={2019},
  publisher={Nature Publishing Group UK London},
  URL={https://doi.org/10.1038/s41565-019-0398-x}
}

@article{bandurin2018fluidity,
  title={Fluidity onset in graphene},
  author={Bandurin, Denis A and Shytov, Andrey V and Levitov, Leonid S and Kumar, Roshan Krishna and Berdyugin, Alexey I and Ben Shalom, Moshe and Grigorieva, Irina V and Geim, Andre K and Falkovich, Gregory},
  journal={Nat. Commun.},
  volume={9},
  number={1},
  pages={4533},
  year={2018},
  publisher={Nature Publishing Group UK London},
  URL={https://doi.org/10.1038/s41467-018-07004-4}
}

@article{doi:10.1126/science.aad0201,
author = {D. A. Bandurin  and I. Torre  and R. Krishna Kumar  and M. Ben Shalom  and A. Tomadin  and A. Principi  and G. H. Auton  and E. Khestanova  and K. S. Novoselov  and I. V. Grigorieva  and L. A. Ponomarenko  and A. K. Geim  and M. Polini },
title = {Negative local resistance caused by viscous electron backflow in graphene},
journal = {Science},
volume = {351},
number = {6277},
pages = {1055-1058},
year = {2016},
doi = {10.1126/science.aad0201}
}

@article{eHD1,
doi = {10.1088/1361-648X/aaa274},
url = {https://dx.doi.org/10.1088/1361-648X/aaa274},
year = {2018},
month = {jan},
publisher = {IOP Publishing},
volume = {30},
number = {5},
pages = {053001},
author = {Andrew Lucas and Kin Chung Fong},
title = {Hydrodynamics of electrons in graphene},
journal = {J. of Phys.: Condens Matter}
}

@article{eHD2,
title = {Electronic hydrodynamics in graphene},
journal = {Ann. Phys.},
volume = {411},
pages = {167979},
year = {2019},
issn = {0003-4916},
doi = {https://doi.org/10.1016/j.aop.2019.167979},
url = {https://www.sciencedirect.com/science/article/pii/S0003491619302349},
author = {Boris N. Narozhny}
}

@Article{eHD3,
author={Narozhny, Boris N.},
title={Hydrodynamic approach to two-dimensional electron systems},
journal={{R}iv. Nuovo Cim.},
year={2022},
month={Oct},
day={01},
volume={45},
number={10},
pages={661-736},
issn={1826-9850},
doi={10.1007/s40766-022-00036-z},
url={https://doi.org/10.1007/s40766-022-00036-z}
}

@article{Aung:2023vrr,
    author = "Aung, Cho Win and Win, Thandar Zaw and Khandal, Gaurav and Ghosh, Sabyasachi",
    title = "{Shear viscosity expression for a graphene system in relaxation time approximation}",
    eprint = "2306.14747",
    archivePrefix = "arXiv",
    primaryClass = "nucl-th",
    doi = "10.1103/PhysRevB.108.235172",
    journal = "Phys. Rev. B",
    volume = "108",
    number = "23",
    pages = "235172",
    year = "2023"
}

@article{Banerjee:2024xnd,
    author = "Banerjee, Soham and Bhadury, Samapan and Florkowski, Wojciech and Jaiswal, Amaresh and Ryblewski, Radoslaw",
    title = "{Longitudinal spin polarization in a thermal model with dissipative corrections}",
    eprint = "2405.05089",
    archivePrefix = "arXiv",
    primaryClass = "hep-ph",
    doi = "10.1103/923l-yxkc",
    journal = "Phys. Rev. C",
    volume = "111",
    number = "6",
    pages = "064912",
    year = "2025"
}

@article{Kapusta:2019sad,
    author = "Kapusta, Joseph I. and Rrapaj, Ermal and Rudaz, Serge",
    title = "{Relaxation Time for Strange Quark Spin in Rotating Quark-Gluon Plasma}",
    eprint = "1907.10750",
    archivePrefix = "arXiv",
    primaryClass = "nucl-th",
    doi = "10.1103/PhysRevC.101.024907",
    journal = "Phys. Rev. C",
    volume = "101",
    number = "2",
    pages = "024907",
    year = "2020"
}

@article{Ayala:2020ndx,
    author = "Ayala, Alejandro and de la Cruz, David and Hern\'andez, L. A. and Salinas, Jordi",
    title = "{Relaxation time for the alignment between the spin of a finite-mass quark or antiquark and the thermal vorticity in relativistic heavy-ion collisions}",
    eprint = "2003.06545",
    archivePrefix = "arXiv",
    primaryClass = "hep-ph",
    doi = "10.1103/PhysRevD.102.056019",
    journal = "Phys. Rev. D",
    volume = "102",
    number = "5",
    pages = "056019",
    year = "2020"
}

@article{Kumar:2023ghs,
    author = {Kumar, Avdhesh and M\"uller, Berndt and Yang, Di-Lun},
    title = "{Spin alignment of vector mesons by glasma fields}",
    eprint = "2304.04181",
    archivePrefix = "arXiv",
    primaryClass = "nucl-th",
    doi = "10.1103/PhysRevD.108.016020",
    journal = "Phys. Rev. D",
    volume = "108",
    number = "1",
    pages = "016020",
    year = "2023"
}

@article{Hidaka:2023oze,
    author = "Hidaka, Yoshimasa and Hongo, Masaru and Stephanov, Mikhail A. and Yee, Ho-Ung",
    title = "{Spin relaxation rate for baryons in a thermal pion gas}",
    eprint = "2312.08266",
    archivePrefix = "arXiv",
    primaryClass = "hep-ph",
    reportNumber = "RIKEN-iTHEMS-Report-23, KEK-TH-2565, J-PARC-TH-0294",
    doi = "10.1103/PhysRevC.109.054909",
    journal = "Phys. Rev. C",
    volume = "109",
    number = "5",
    pages = "054909",
    year = "2024"
}

@article{Wagner:2024fhf,
    author = "Wagner, David and Shokri, Masoud and Rischke, Dirk H.",
    title = "{Damping of spin waves}",
    eprint = "2405.00533",
    archivePrefix = "arXiv",
    primaryClass = "nucl-th",
    doi = "10.1103/PhysRevResearch.6.043103",
    journal = "Phys. Rev. Res.",
    volume = "6",
    number = "4",
    pages = "043103",
    year = "2024"
}

@article{Becattini:2014yxa,
    author = "Becattini, F. and Bucciantini, L. and Grossi, E. and Tinti, L.",
    title = "{Local thermodynamical equilibrium and the beta frame for a quantum relativistic fluid}",
    eprint = "1403.6265",
    archivePrefix = "arXiv",
    primaryClass = "hep-th",
    doi = "10.1140/epjc/s10052-015-3384-y",
    journal = "Eur. Phys. J. C",
    volume = "75",
    number = "5",
    pages = "191",
    year = "2015"
}

@article{Becattini:2018duy,
    author = "Becattini, F. and Florkowski, Wojciech and Speranza, Enrico",
    title = "{Spin tensor and its role in non-equilibrium thermodynamics}",
    eprint = "1807.10994",
    archivePrefix = "arXiv",
    primaryClass = "hep-th",
    doi = "10.1016/j.physletb.2018.12.016",
    journal = "Phys. Lett. B",
    volume = "789",
    pages = "419--425",
    year = "2019"
}

@article{Becattini:2017gcx,
    author = "Becattini, F. and Karpenko, Iu.",
    title = "{Collective Longitudinal Polarization in Relativistic Heavy-Ion Collisions at Very High Energy}",
    eprint = "1707.07984",
    archivePrefix = "arXiv",
    primaryClass = "nucl-th",
    doi = "10.1103/PhysRevLett.120.012302",
    journal = "Phys. Rev. Lett.",
    volume = "120",
    number = "1",
    pages = "012302",
    year = "2018"
}

@article{Florkowski:2019voj,
    author = "Florkowski, Wojciech and Kumar, Avdhesh and Ryblewski, Radoslaw and Mazeliauskas, Aleksas",
    title = "{Longitudinal spin polarization in a thermal model}",
    eprint = "1904.00002",
    archivePrefix = "arXiv",
    primaryClass = "nucl-th",
    doi = "10.1103/PhysRevC.100.054907",
    journal = "Phys. Rev. C",
    volume = "100",
    number = "5",
    pages = "054907",
    year = "2019"
}

@article{Karpenko:2016jyx,
    author = "Karpenko, I. and Becattini, F.",
    title = "{Study of $\Lambda $ polarization in relativistic nuclear collisions at $\sqrt{s_\mathrm {NN}}=7.7$ \textendash{}200 GeV}",
    eprint = "1610.04717",
    archivePrefix = "arXiv",
    primaryClass = "nucl-th",
    doi = "10.1140/epjc/s10052-017-4765-1",
    journal = "Eur. Phys. J. C",
    volume = "77",
    number = "4",
    pages = "213",
    year = "2017"
}

@article{Bhadury:2022ulr,
    author = "Bhadury, Samapan and Florkowski, Wojciech and Jaiswal, Amaresh and Kumar, Avdhesh and Ryblewski, Radoslaw",
    title = "{Relativistic Spin Magnetohydrodynamics}",
    eprint = "2204.01357",
    archivePrefix = "arXiv",
    primaryClass = "nucl-th",
    doi = "10.1103/PhysRevLett.129.192301",
    journal = "Phys. Rev. Lett.",
    volume = "129",
    number = "19",
    pages = "192301",
    year = "2022"
}

@article{Hehl:1976vr,
    author = "Hehl, F. W.",
    title = "{On the Energy Tensor of Spinning Massive Matter in Classical Field Theory and General Relativity}",
    doi = "10.1016/0034-4877(76)90016-1",
    journal = "Rept. Math. Phys.",
    volume = "9",
    pages = "55--82",
    year = "1976"
}

@article{Wallace:1947qeg,
    author = "Wallace, P. R.",
    title = "{The Band Theory of Graphite}",
    doi = "10.1103/PhysRev.71.622",
    journal = "Phys. Rev.",
    volume = "71",
    number = "9",
    pages = "622",
    year = "1947"
}

@article{Novoselov:2005kj,
    author = "Novoselov, K. S. and Geim, A. K. and Morozov, S. V. and Jiang, D. and Katsnelson, M. I. and Grigorieva, I. V. and Dubonos, S. V. and Firsov, A. A.",
    title = "{Two-dimensional gas of massless Dirac fermions in graphene}",
    eprint = "cond-mat/0509330",
    archivePrefix = "arXiv",
    doi = "10.1038/nature04233",
    journal = "Nature",
    volume = "438",
    pages = "197",
    year = "2005"
}

@article{Zhang:2005zz,
    author = "Zhang, Yuanbo and Tan, Yan-Wen and Stormer, Horst L. and Kim, Philip",
    title = "{Experimental observation of the quantum Hall effect and and Berry's phase in graphene}",
    eprint = "cond-mat/0509355",
    archivePrefix = "arXiv",
    doi = "10.1038/nature04235",
    journal = "Nature",
    volume = "438",
    pages = "201--204",
    year = "2005"
}

@article{Barnett:1935wyv,
    author = "Barnett, S. J.",
    title = "{Gyromagnetic and Electron-Inertia Effects}",
    doi = "10.1103/RevModPhys.7.129",
    journal = "Rev. Mod. Phys.",
    volume = "7",
    number = "2",
    pages = "129",
    year = "1935"
}

@article{Idrisov:2023,
    author = {Idrisov, Edvin G. and Hasdeo, Eddwi H. and Radhakrishnan, Byjesh N. and Schmidt, Thomas L.},
    title = "{Hydrodynamic Navier-Stokes equations in two-dimensional systems with Rashba spin-orbit coupling}",
    journal = {Low Temperature Physics},
    volume = {49},
    number = {12},
    pages = {1385-1397},
    year = {2023},
    month = {12},
    issn = {1063-777X},
    doi = {10.1063/10.0022364},
    url = {https://doi.org/10.1063/10.0022364},
}

@article{PhysRevB.84.054437,
  title = {Joule heating in nanowires},
  author = {Fangohr, Hans and Chernyshenko, Dmitri S. and Franchin, Matteo and Fischbacher, Thomas and Meier, Guido},
  journal = {Phys. Rev. B},
  volume = {84},
  issue = {5},
  pages = {054437},
  numpages = {12},
  year = {2011},
  month = {Aug},
  publisher = {American Physical Society},
  doi = {10.1103/PhysRevB.84.054437},
  url = {https://link.aps.org/doi/10.1103/PhysRevB.84.054437}
}

@article{PhysRevLett.98.187202,
  title = {Direct Imaging of Stochastic Domain-Wall Motion Driven by Nanosecond Current Pulses},
  author = {Meier, Guido and Bolte, Markus and Eiselt, Ren\'e and Kr\"uger, Benjamin and Kim, Dong-Hyun and Fischer, Peter},
  journal = {Phys. Rev. Lett.},
  volume = {98},
  issue = {18},
  pages = {187202},
  numpages = {4},
  year = {2007},
  month = {May},
  publisher = {American Physical Society},
  doi = {10.1103/PhysRevLett.98.187202},
  url = {https://link.aps.org/doi/10.1103/PhysRevLett.98.187202}
}

@article{10.1063/1.1847714,
    author = {Yamaguchi, A. and Nasu, S. and Tanigawa, H. and Ono, T. and Miyake, K. and Mibu, K. and Shinjo, T.},
    title = {Effect of Joule heating in current-driven domain wall motion},
    journal = {Applied Physics Letters},
    volume = {86},
    number = {1},
    pages = {012511},
    year = {2004},
    month = {12},
    issn = {0003-6951},
    doi = {10.1063/1.1847714},
    url = {https://doi.org/10.1063/1.1847714}
}

@article{Crossno2016,
author = {Jesse Crossno  and Jing K. Shi  and Ke Wang  and Xiaomeng Liu  and Achim Harzheim  and Andrew Lucas  and Subir Sachdev  and Philip Kim  and Takashi Taniguchi  and Kenji Watanabe  and Thomas A. Ohki  and Kin Chung Fong },
title = {Observation of the Dirac fluid and the breakdown of the Wiedemann-Franz law in graphene},
journal = {Science},
volume = {351},
number = {6277},
pages = {1058-1061},
year = {2016},
doi = {10.1126/science.aad0343},
URL = {https://www.science.org/doi/abs/10.1126/science.aad0343}
}

@article{Lucas2016,
  title = {Transport in inhomogeneous quantum critical fluids and in the Dirac fluid in graphene},
  author = {Lucas, Andrew and Crossno, Jesse and Fong, Kin Chung and Kim, Philip and Sachdev, Subir},
  journal = {Phys. Rev. B},
  volume = {93},
  issue = {7},
  pages = {075426},
  numpages = {17},
  year = {2016},
  month = {Feb},
  publisher = {American Physical Society},
  doi = {10.1103/PhysRevB.93.075426},
  url = {https://link.aps.org/doi/10.1103/PhysRevB.93.075426}
}

@article{Gallagher2019,
author = {Patrick Gallagher  and Chan-Shan Yang  and Tairu Lyu  and Fanglin Tian  and Rai Kou  and Hai Zhang  and Kenji Watanabe  and Takashi Taniguchi  and Feng Wang },
title = {Quantum-critical conductivity of the Dirac fluid in graphene},
journal = {Science},
volume = {364},
number = {6436},
pages = {158-162},
year = {2019},
doi = {10.1126/science.aat8687},
URL = {https://www.science.org/doi/abs/10.1126/science.aat8687}
}

@article{Lucas2019,
author = {Andrew Lucas },
title = {An exotic quantum fluid in graphene},
journal = {Science},
volume = {364},
number = {6436},
pages = {125-125},
year = {2019},
doi = {10.1126/science.aaw9869},
URL = {https://www.science.org/doi/abs/10.1126/science.aaw9869}
}

@article{Burkov:2017rgl,
    author = "Burkov, A. A.",
    title = "{Weyl Metals}",
    eprint = "1704.06660",
    archivePrefix = "arXiv",
    primaryClass = "cond-mat.mes-hall",
    doi = "10.1146/annurev-conmatphys-033117-054129",
    journal = "Ann. Rev. Condensed Matter Phys.",
    volume = "9",
    pages = "359--378",
    year = "2018"
}

@article{Vafek:2013mpa,
    author = "Vafek, Oskar and Vishwanath, Ashvin",
    title = "{Dirac Fermions in Solids: From High-Tc cuprates and Graphene to Topological Insulators and Weyl Semimetals}",
    eprint = "1306.2272",
    archivePrefix = "arXiv",
    primaryClass = "cond-mat.mes-hall",
    doi = "10.1146/annurev-conmatphys-031113-133841",
    journal = "Ann. Rev. Condensed Matter Phys.",
    volume = "5",
    pages = "83--112",
    year = "2014"
}

@article{Armitage:2017cjs,
    author = "Armitage, N. P. and Mele, E. J. and Vishwanath, Ashvin",
    title = "{Weyl and Dirac Semimetals in Three Dimensional Solids}",
    eprint = "1705.01111",
    archivePrefix = "arXiv",
    primaryClass = "cond-mat.str-el",
    doi = "10.1103/RevModPhys.90.015001",
    journal = "Rev. Mod. Phys.",
    volume = "90",
    number = "1",
    pages = "015001",
    year = "2018"
}

@article{Yan:2016euz,
    author = "Yan, Binghai and Felser, Claudia",
    title = "{Topological Materials: Weyl Semimetals}",
    eprint = "1611.04182",
    archivePrefix = "arXiv",
    primaryClass = "cond-mat.mtrl-sci",
    doi = "10.1146/annurev-conmatphys-031016-025458",
    journal = "Ann. Rev. Condensed Matter Phys.",
    volume = "8",
    pages = "337--354",
    year = "2017"
}

@article{Hosur:2013kxa,
    author = "Hosur, Pavan and Qi, Xiaoliang",
    title = "{Recent developments in transport phenomena in Weyl semimetals}",
    eprint = "1309.4464",
    archivePrefix = "arXiv",
    primaryClass = "cond-mat.str-el",
    doi = "10.1016/j.crhy.2013.10.010",
    journal = "Comptes Rendus Physique",
    volume = "14",
    pages = "857--870",
    year = "2013"
}

@article{RAVICHANDRAN201514,
title = {Transient Joule heating of graphene, nanowires and filaments: Analytical model for current-induced temperature evolution including substrate and end effects},
journal = {International Journal of Heat and Mass Transfer},
volume = {88},
pages = {14-19},
year = {2015},
issn = {0017-9310},
doi = {https://doi.org/10.1016/j.ijheatmasstransfer.2015.04.014},
url = {https://www.sciencedirect.com/science/article/pii/S0017931015003749},
author = {K.S. {Ravi Chandran}},
}

@article{Becattini:2011ev,
    author = "Becattini, F. and Tinti, L.",
    title = "{Thermodynamical inequivalence of quantum stress-energy and spin tensors}",
    eprint = "1101.5251",
    archivePrefix = "arXiv",
    primaryClass = "hep-th",
    doi = "10.1103/PhysRevD.84.025013",
    journal = "Phys. Rev. D",
    volume = "84",
    pages = "025013",
    year = "2011"
}

@article{Becattini:2025twu,
    author = "Becattini, Francesco and Hoyos, Carlos",
    title = "{Pseudo-gauge invariant non-equilibrium density operator}",
    eprint = "2507.09249",
    archivePrefix = "arXiv",
    primaryClass = "nucl-th",
    doi = "10.1103/sk54-wd8z",
    journal = "Phys. Rev. Lett.",
    volume = "137",
    number = "4",
    pages = "041602",
    year = "2026"
}

@article{10.1063/5.0259803,
    author = {Wang, Kang and Shi, Meng and Li, Xueqin and Wang, Ning and Qu, Zhe and Du, Haifeng},
    title = {Compact dynamic cantilever magnetometry},
    journal = {Review of Scientific Instruments},
    volume = {96},
    number = {4},
    pages = {045104},
    year = {2025},
    month = {04},
    issn = {0034-6748},
    doi = {10.1063/5.0259803},
    url = {https://doi.org/10.1063/5.0259803},
}

@article{
doi:10.1073/pnas.0914031107,
author = {Chi Hwan Lee  and Dong Rip Kim  and Xiaolin Zheng },
title = {Fabricating nanowire devices on diverse substrates by simple transfer-printing methods},
journal = {Proceedings of the National Academy of Sciences},
volume = {107},
number = {22},
pages = {9950-9955},
year = {2010},
doi = {10.1073/pnas.0914031107},
URL = {https://www.pnas.org/doi/abs/10.1073/pnas.0914031107},
}

@article{Xu_2012,
doi = {10.1088/0953-8984/24/47/475305},
url = {https://dx.doi.org/10.1088/0953-8984/24/47/475305},
year = {2012},
month = {nov},
publisher = {IOP Publishing},
volume = {24},
number = {47},
pages = {475305},
author = {Xu, Zhiping and Buehler, Markus J},
title = {Heat dissipation at a graphene–substrate interface},
journal = {Journal of Physics: Condensed Matter},
}

@article{PhysRevLett.112.116602,
  title = {Spin Relaxation Mechanism in Graphene: Resonant Scattering by Magnetic Impurities},
  author = {Kochan, Denis and Gmitra, Martin and Fabian, Jaroslav},
  journal = {Phys. Rev. Lett.},
  volume = {112},
  issue = {11},
  pages = {116602},
  numpages = {5},
  year = {2014},
  month = {Mar},
  publisher = {American Physical Society},
  doi = {10.1103/PhysRevLett.112.116602},
  url = {https://link.aps.org/doi/10.1103/PhysRevLett.112.116602}
}

@article{HAN2012369,
title = {Spin transport and relaxation in graphene},
journal = {Journal of Magnetism and Magnetic Materials},
volume = {324},
number = {4},
pages = {369-381},
year = {2012},
issn = {0304-8853},
doi = {https://doi.org/10.1016/j.jmmm.2011.08.001},
url = {https://www.sciencedirect.com/science/article/pii/S0304885311005373},
author = {Wei Han and K.M. McCreary and K. Pi and W.H. Wang and Yan Li and H. Wen and J.R. Chen and R.K. Kawakami}
}

@article{VanTuan2016,
author={Van Tuan, Dinh
and Ortmann, Frank
and Cummings, Aron W.
and Soriano, David
and Roche, Stephan},
title={Spin dynamics and relaxation in graphene dictated by electron-hole puddles},
journal={Scientific Reports},
year={2016},
month={Feb},
day={15},
volume={6},
number={1},
pages={21046},
issn={2045-2322},
doi={10.1038/srep21046},
url={https://doi.org/10.1038/srep21046}
}

@article{RevModPhys.81.109,
  title = {The electronic properties of graphene},
  author = {Castro Neto, A. H. and Guinea, F. and Peres, N. M. R. and Novoselov, K. S. and Geim, A. K.},
  journal = {Rev. Mod. Phys.},
  volume = {81},
  issue = {1},
  pages = {109--162},
  numpages = {0},
  year = {2009},
  month = {Jan},
  publisher = {American Physical Society},
  doi = {10.1103/RevModPhys.81.109},
  url = {https://link.aps.org/doi/10.1103/RevModPhys.81.109}
}

@article{Tinti:2016bav,
    author = "Tinti, Leonardo and Jaiswal, Amaresh and Ryblewski, Radoslaw",
    title = "{Quasiparticle second-order viscous hydrodynamics from kinetic theory}",
    eprint = "1612.07329",
    archivePrefix = "arXiv",
    primaryClass = "nucl-th",
    doi = "10.1103/PhysRevD.95.054007",
    journal = "Phys. Rev. D",
    volume = "95",
    number = "5",
    pages = "054007",
    year = "2017"
}

@article{Jaiswal:2012qm,
    author = "Jaiswal, Amaresh and Bhalerao, Rajeev S. and Pal, Subrata",
    title = "{New relativistic dissipative fluid dynamics from kinetic theory}",
    eprint = "1204.3779",
    archivePrefix = "arXiv",
    primaryClass = "nucl-th",
    reportNumber = "TIFR-TH-12-14",
    doi = "10.1016/j.physletb.2013.02.025",
    journal = "Phys. Lett. B",
    volume = "720",
    pages = "347--351",
    year = "2013"
}

@article{RevModPhys.87.1213,
  title = {Spin Hall effects},
  author = {Sinova, Jairo and Valenzuela, Sergio O. and Wunderlich, J. and Back, C. H. and Jungwirth, T.},
  journal = {Rev. Mod. Phys.},
  volume = {87},
  issue = {4},
  pages = {1213--1260},
  numpages = {47},
  year = {2015},
  month = {Oct},
  publisher = {American Physical Society},
  doi = {10.1103/RevModPhys.87.1213},
  url = {https://link.aps.org/doi/10.1103/RevModPhys.87.1213}
}

@article{PhysRevLett.95.226801,
  title = {Quantum Spin Hall Effect in Graphene},
  author = {Kane, C. L. and Mele, E. J.},
  journal = {Phys. Rev. Lett.},
  volume = {95},
  issue = {22},
  pages = {226801},
  numpages = {4},
  year = {2005},
  month = {Nov},
  publisher = {American Physical Society},
  doi = {10.1103/PhysRevLett.95.226801},
  url = {https://link.aps.org/doi/10.1103/PhysRevLett.95.226801}
}
%%

%%%%%%%%%%%%%%%%%%%%%%%%%%%%%%%%%%%%%%%%%%%%%%%%%%%%%%%%%%%%%%%%%%%%%%%%%

\end{document}